\documentclass[prc,twocolumn,showpacs,amsmath,amssymb,superscriptaddress,floatfix,nofootinbib,10pt]{revtex4}

\usepackage{slashed}
\usepackage{mathrsfs,bm}
\usepackage{longtable,lscape}
\usepackage{txfonts}
\usepackage{amssymb}
\usepackage{indentfirst}
\usepackage{graphicx,booktabs}
\usepackage{multirow}
\usepackage{overpic}
\usepackage{color}
\usepackage{amssymb}
\usepackage[utf8]{inputenc}

\begin{document}
\title{$D \Xi$ and $D^* \Xi$
  Molecular States from One Boson Exchange}

\author{Ming-Zhu Liu}
\affiliation{School of Physics and
Nuclear Energy Engineering \& International Research Center for Nuclei and Particles in the Cosmos \&
Beijing Key Laboratory of Advanced Nuclear Materials and Physics,  Beihang University, Beijing 100191, China}

\author{Tian-Wei Wu}
\affiliation{School of Physics and
Nuclear Energy Engineering \& International Research Center for Nuclei and Particles in the Cosmos \&
Beijing Key Laboratory of Advanced Nuclear Materials and Physics,  Beihang University, Beijing 100191, China}

\author{Ju-Jun Xie}
\affiliation{Institute of Modern Physics, Chinese Academy of
  Sciences, Lanzhou 730000, China}

\author{Manuel Pavon Valderrama}
\email[E-mail: ]{mpavon@buaa.edu.cn}
\affiliation{School of Physics and
Nuclear Energy Engineering \& International Research Center for Nuclei and Particles in the Cosmos \&
Beijing Key Laboratory of Advanced Nuclear Materials and Physics,  Beihang University, Beijing 100191, China}

\author{Li-Sheng Geng}
\email[E-mail: ]{lisheng.geng@buaa.edu.cn}
\affiliation{School of Physics and
Nuclear Energy Engineering \& International Research Center for Nuclei and Particles in the Cosmos \&
Beijing Key Laboratory of Advanced Nuclear Materials and Physics,  Beihang University, Beijing 100191, China}
\date{\today}
\begin{abstract}
  We explore the existence of $D \Xi$ and $D^* \Xi$ molecular states
  within the one boson exchange model.
  We regularize the potential derived in this model with a form factor
  and a cut-off of the order of $1\,{\rm GeV}$.
  To determine the cut-off, we use the condition that
  the $X(3872)$ is reproduced as a pole
  in the $J^{PC} = 1^{++}$ $D^*\bar{D}$ amplitude.
  From this we find that the $J^P= {\frac{1}{2}}^{-}$ $D^*\,\Xi$ system
  is on the verge of binding and has an unnaturally large scattering length.
  For the $J^P= {\frac{1}{2}}^{-}$ $D\,\Xi$ and
  the $J^P= {\frac{3}{2}}^{-}$ $D^*\,\Xi$ systems
  the attraction is not enough to form a bound state.  
  From heavy quark symmetry and the quark model
  we can extend the previous model to the $P \Xi_{QQ}$ and $P^* \Xi_{QQ}$ systems,
  with $P = B, D$ and $\Xi_{QQ} = \Xi_{cc}, \Xi_{bb}$.
  In this case we predict a series of triply heavy pentaquark-like molecules.
\end{abstract}


\maketitle
\section{INTRODUCTION}

The discovery of the $X(3872)$ by the Belle Collaboration~\cite{Choi:2003ue}
fifteen years ago represented the first hidden charm state
that did not fit into the charmonium spectrum.
Afterwards experiments have found a series of similar states, informally
known as XYZ states.
They cannot be easily accommodated in the naive quark model and
other components have to be invoked to explain their masses,
decay widths and  production rates, see~\cite{Guo:2017jvc}
for a recent review.
A few are tetraquark-like, such as
the $Z_{c}(3900)$~\cite{Ablikim:2013mio, Liu:2013dau},
$Z_{c}(4020)$~\cite{Ablikim:2013wzq, Ablikim:2014dxl},
$Z_{b}(10610)$ and $Z_{b}(10650)$~\cite{Belle:2011aa,Garmash:2014dhx},
while recently two pentaquark-like states have been observed,
the $P_{c}(4380)$ and $P_{c}(4450)$~\cite{Aaij:2015tga}.
A few of them have the interesting feature of
being close to an open flavour threshold.
The most notable example is the $X(3872)$, located almost on top of
the $D^0\bar{D}^{0*}$ threshold, but the list also includes
the $Z_{c}(3900)$ ($D\bar{D}^{\ast}$) and $Z_{c}(4020)$ ($D^{\ast}\bar{D}^{\ast}$),
the $Z_{b}(10610)$ ($B\bar{B}^{\ast}$) and $Z_{b}(10650)$ ($B^{\ast}\bar{B}^{\ast}$)
and the $P_{c}(4450)$ near the $\bar{D}^{\ast}\Sigma_{c}$ threshold.
This characteristic has led to the conjecture that
the previous states might be hadronic molecules.

A hadronic molecule is a loosely bound state or resonance composed of hadrons.
They were originally proposed to explain the $\psi(4040)$ as a
$D^{\ast}\bar{D}^{\ast}$ bound state~\cite{Voloshin:1976ap,DeRujula:1976zlg}.
Though the $\psi(4040)$ turned out to be a charmonium state at the end,
the idea quickly caught the attention of theoreticians~\cite{Tornqvist:1991ks,Tornqvist:1993ng,Manohar:1992nd,Ericson:1993wy} and
the later discovery of the $X(3872)$ showed that these speculations
were indeed on the right track.
Besides the $X(3872)$, the most prosaic example of a hadronic molecule
is the deuteron, which also inspired the Weinberg compositeness
condition \cite{Weinberg:1965zz}.
Other strong molecular candidates include the $Z_{b}(10610)$, $Z_{b}(10650)$~\cite{Liu:2008fh, Liu:2008tn, Sun:2011uh, He:2015mja, Sun:2012zzd,Dias:2014pva,Wang:2018jlv}
and the $P_c(4450)$~\cite{Chen:2015loa,Chen:2015moa,Roca:2015dva,He:2015cea,Xiao:2015fia}.

Recently the LHCb Collaboration has observed five narrow states
$\Omega_{c}(3000)^{0}$, $\Omega_{c}(3050)^{0}$, $\Omega_{c}(3066)^{0}$, $\Omega_{c}(3090)^{0}$, and $\Omega_{c}(3119)^{0}$~\cite{Aaij:2017nav},
where four of them have also been recently confirmed
from an analysis of the Belle data~\cite{Yelton:2017qxg}.
These states can be accomodated as excited $\Omega_c$ baryons~\cite{Cheng:2017ove,Wang:2017hej}, as compact baryons in which the $ss$ strange quark pair
forms a diquark~\cite{Karliner:2017kfm} or as molecular states~\cite{Debastiani:2017ewu,Chen:2017xat,Montana:2017kjw,Nieves:2017jjx,Huang:2018wgr}.
The LHCb data also hint at a structure around $3188\,{\rm MeV}$,
which is near the $D\Xi$ threshold ($3179-3191\,{\rm MeV}$),
and could be a bound state of $D\Xi$.
Wang et al.~\cite{Wang:2017smo} used the Bethe-Salpeter formalism
to study the  $S-$wave $D\Xi$ interaction and
they found two bound states, one isoscalar and one isovector,
respectively, which could contribute to the $3188\,{\rm MeV}$ structure
near the five new narrow $\Omega_{c}$ states.
Debastiani and Liang~\cite{Debastiani:2017ewu,Liang:2017ejq}
used an extension of the chiral unitary model to calculate
the interactions between $D\Xi$, $D^{\ast}\Xi$ ,$\bar{B}\Xi$, $\bar{B}^{\ast}\Xi$
plus other channels, and obtained  one zero width state
with $J^{P}=1/2^{-},3/2^{-}$ ,
which couples mostly to $D^{\ast}\Xi$ and $\bar{B}^{\ast}\Xi$
and another state with $J^{P}=\frac{1}{2}^{-}$,
which couples mostly to $D\Xi$ and $\bar{B}\Xi$.

In this work we study possible bound states near the thresholds of
$D\Xi$, $D^{\ast}\Xi$, $\bar{B}\Xi$ and $\bar{B}^{\ast}\Xi$
within the one boson exchange (OBE) model,
where we will also consider the replacement of $\Xi$ by $\Xi_{cc}$ or $\Xi_{bb}$
to explore the existence of heavy molecules containing
two and three charm/bottom quarks.
The OBE model is an intuitive framework in which
a few of the most quantitative succesful descriptions of
the nuclear force have been achieved~\cite{Machleidt:1987hj,Machleidt:1989tm}.
Nowadays it has been superseded by the effective field theory (EFT)
approaches~\cite{Bedaque:2002mn,Machleidt:2011zz},
which have two theoretical advantages over the OBE model:
(i) the possibility of making {\it a priori} error estimates and
(ii) the low-energy equivalence with quantum chromodynamics.
However the EFT approach requires a large number of data to fix
the low energy constants that substitute the exchange of
light mesons such as the $\sigma$, $\rho$ and $\omega$.
This means that in situations where hadron-hadron scattering data is poor,
such as hadron molecules, the OBE potential has the advantage of
providing a model of the short-range dynamics of these systems,
at the price of sacrificing the systematicity of EFT.
As a matter of fact the seminal works that pioneered the idea of
hadronic molecules~\cite{Voloshin:1976ap,DeRujula:1976zlg}
were indeed based on the OBE model and a few modern explorations
rely heavily on this model~\cite{Liu:2008tn,Sun:2011uh,Chen:2015loa}.

The manuscript is organized as follows:
in Section \ref{sec:OBE} we review the OBE model as applied to the $D \Xi$,
$D^* \Xi$ plus analogous systems and derive the potentials in these systems.
In Section \ref{sec:predictions} we show the predictions we obtain
for the molecular states.
Finally we present our conclusions in Section \ref{sec:summary}.

\section{The One Boson Exchange Potential}
\label{sec:OBE}

In this section we explain the one boson exchange (OBE) model
(see Ref.~\cite{Machleidt:1989tm} for a review) and
how it applies to the $D \Xi$ and $D^{\ast}\Xi$ systems.
The OBE model is a generalization of the idea of Yukawa,
namely that nuclear forces arise from the exchange of pions,
to shorter distances.
For that the exchange of other light mesons (besides the pion)
is considered, in particular the $\sigma$ scalar meson
and the $\rho$ and $\omega$ vector mesons.
The reason for the development of the OBE model was the failure of
the original two-pion theories in the fifties (which
did not include chiral symmetry), see Ref.~\cite{Machleidt:2017vls}
for a historical perspective.
The fundamental idea is that the bulk of the two-pion exchange potential
can be described by the exchange of a heavier meson, which either couples
strongly to pions or which can be interpreted as a  multi-pion resonance.
The idea is physically compelling and has been fairly successful
at the phenomenological level, as illustrated
by the nuclear potentials based on it~\cite{Machleidt:1987hj,Machleidt:2000ge}.
There are limitations however, such as the requirement of form factors
to regularize the potentials or the requirement of fine-tuning
the coupling constants~\cite{CalleCordon:2008cz,Cordon:2009pj}
(at least in the two-nucleon system).
Yet the OBE potential is still perfectly able to provide a good estimation
of the plausability of hadronic bound states and their location,
as attested for instance in the pioneering work
of Voloshin and Okun~\cite{Voloshin:1976ap}.

\subsection{The Lagrangian}

We begin with the interaction of the $D$ and $D^*$ fields
with the $\pi$, $\sigma$, $\rho$ and $\omega$ mesons.
First we group the fields of the $D$ and $D^*$
in the (heavy spin symmetric) superfield
\begin{eqnarray}
  H_v = \frac{1 + \slashed v}{2 \sqrt{2}} \left(
  D^{*\mu} \gamma_{\mu} - D \gamma_5
  \right) \, ,
\end{eqnarray}
with $v$ the velocity parameter and where we have used
the convenient normalization of Falk and Luke~\cite{Falk:1992cx}.
In this normalization we can write the interaction lagrangian of
the $H$ superfield with the light mesons as
\begin{eqnarray}
  \mathcal{L}_{H H \pi} &=& \frac{g_1}{\sqrt{2} f_{\pi}}\,{\rm Tr}
  \left[ \bar{H}_v \gamma^{\mu} \gamma^5 \vec{\tau} \cdot
    \partial_{\mu} \vec{\pi} H_v \right] \, , \\
  \mathcal{L}_{H H \sigma} &=& -g_{\sigma 1}\,{\rm Tr}\left[
    \bar{H}_v \sigma H_v \right] \, , \\
  \mathcal{L}_{H H \rho} &=& g_{\rho 1}\,{\rm Tr}\left[
    \bar{H}_v v_{\mu} \vec{\tau} \cdot \vec{\rho}^{\mu} H_v \right]
  \nonumber \\
  &+& \frac{f_{\rho 1}}{4 M_1}\,{\rm Tr}\left[
    \bar{H}_v \sigma_{\mu \nu} \vec{\tau} \cdot \left( \partial^\mu \vec{\rho}^{\nu}
    - \partial^\nu \vec{\rho}^{\mu} \right) H_v \right] \, , \\
    \mathcal{L}_{H H \omega} &=& -g_{\omega 1}\,{\rm Tr}\left[
    \bar{H}_v v_{\mu} {\omega}^{\mu} H_v \right]
  \nonumber \\
  &-& \frac{f_{\rho 1}}{4 M_1}\,{\rm Tr}\left[
    \bar{H}_v \sigma_{\mu \nu} \vec{\tau} \, \left( \partial^\mu {\omega}^{\nu}
    - \partial^\nu {\omega}^{\mu} \right) H_v \right] \, ,
\end{eqnarray}
where the traces are in the 4x4 space spanned by the Dirac matrices
when defining the superfield $H$.
In the lagrangian above we take the normalization $f_{\pi} = 132\,{\rm MeV}$,
$g_1$, $g_{\sigma 1}$, $g_{\omega 1}$, $f_{\omega 1}$, $g_{\rho 1}$ and $f_{\rho 1}$
are the different couplings in the OBE model and $M_1$ is a mass scale
that we introduce for $f_{\omega 1}$ and $f_{\rho 1}$ to be dimensionless
($M_1$ is in principle arbitrary, but we will later take it to be the D-meson
mass, i.e. $M_1 = m_D$).
If we choose the velocity parameter to be $v = (1,\vec{0})$,
we can substitute the superfield $H_v$ by a non-relativistic superfield $H$
\begin{eqnarray}
  H_v \to H = \frac{1}{\sqrt{2}}\,
  \left[ P + \vec{P}^* \cdot \vec{\sigma} \right] \, ,
\end{eqnarray}
where $H$ is a 2x2 matrix (instead of a 4x4 one).
Now the lagrangian can be rewritten as
\begin{eqnarray}
    \mathcal{L}_{H H \pi} &=& -\frac{g_1}{\sqrt{2} f_{\pi}}\,{\rm Tr}
  \left[ {H}^{\dagger} \vec{\sigma} \cdot \nabla ( \vec{\tau} \cdot \vec{\pi})
    H \right] \, , \\
  \mathcal{L}_{H H \sigma} &=& g_{\sigma 1}\,{\rm Tr}\left[
    {H}^{\dagger} \sigma H \right] \, , \\
  \mathcal{L}_{H H \rho} &=& g_{\rho 1}\,{\rm Tr}\left[
    {H}^{\dagger} \vec{\tau} \cdot \vec{\rho}^{0} H \right]
  \nonumber \\
  &-& \frac{f_{\rho 1}}{4 M_1}\,\epsilon_{ijk}\,{\rm Tr}\left[
    {H}^{\dagger} \sigma_k \vec{\tau} \cdot \left( \partial_i \vec{\rho}_j
    - \partial_j \vec{\rho}_i \right) H \right] \, , \\
    \mathcal{L}_{H H \omega} &=& -g_{\omega 1}\,{\rm Tr}\left[
    {H}^{\dagger} {\omega}^{0} H \right]
  \nonumber \\
  &+& \frac{f_{\omega 1}}{4 M_1}\,\epsilon_{ijk}\,{\rm Tr}\left[
    {H}^{\dagger} \sigma_{k} \, \left( \partial_i {\omega}_j
    - \partial_j {\omega}_i \right) H \right] \, .
\end{eqnarray}

Now for the $\Xi$ field we can write the interaction lagrangian
\begin{eqnarray}
  \mathcal{L}_{\Xi \Xi \pi} &=& -
  \frac{g_2}{\sqrt{2} f_{\pi}}\,\bar{\psi}_{\Xi} \gamma^{\mu} \gamma^5 \vec{\tau} \cdot
  \partial_{\mu} \vec{\pi} \psi_{\Xi} \, , \\
  \mathcal{L}_{\Xi \Xi \sigma} &=& g_{\sigma 2}
  \,\bar{\psi}_{\Xi} \sigma \psi_{\Xi} \, , \\
  \mathcal{L}_{\Xi \Xi \rho} &=& g_{\rho 2}
  \,\bar{\psi}_{\Xi} \gamma_\mu \vec{\tau} \cdot \vec{\rho}^{\mu}  \psi_{\Xi} \nonumber \\
  &+& \frac{f_{\rho 2}}{4 M_2} \,\bar{\psi}_{\Xi} \sigma_{\mu,\nu}
  \vec{\tau} \cdot \left( \partial^\mu \vec{\rho}^{\nu} -
  \partial^{\nu} \vec{\rho}^{\mu}
  \right)  \psi_{\Xi} \, , \\
  \mathcal{L}_{\Xi \Xi \omega} &=& g_{\omega 2}
  \,\bar{\psi}_{\Xi} \gamma_\mu {\omega}^{\mu}  \psi_{\Xi} \nonumber \\
  &+& \frac{f_{\omega 2}}{4 M_2} \,\bar{\psi}_{\Xi} \sigma_{\mu,\nu}
  \,\left( \partial^\mu {\omega}^{\nu} - \partial^{\nu} {\omega}^{\mu}
  \right)  \psi_{\Xi} \, ,
\end{eqnarray}
which is analogous to the one for the $D$ and $D^*$ (and it is identical
in form to the one in the nucleon-nucleon case).
In the lagrangian $g_2$, $g_{\sigma 2}$, $g_{\omega 2}$, $f_{\omega 2}$, $g_{\rho 2}$,
$f_{\rho 2}$ and $M_2$ denote the couplings and the mass scale
for the cascade baryon.
Here we can use the heavy baryon formulation by making the field redefinition
\begin{eqnarray}
  \Xi = e^{i M_{\Xi} v \cdot x}\,\psi_{\Xi} \, .
\end{eqnarray}
If we choose again $v = (1,\vec{0})$, we arrive at the lagrangian
\begin{eqnarray}
  \mathcal{L}_{\Xi \Xi \pi} &=&
  \frac{g_2}{\sqrt{2} f_{\pi}}\,
       {\Xi}^{\dagger}
       \vec{\sigma} \cdot \nabla ( \vec{\tau} \cdot \vec{\pi})  
       \Xi \, , \\
  \mathcal{L}_{\Xi \Xi \sigma} &=& g_{\sigma 2}
  \,\Xi^{\dagger} \sigma \Xi \, , \\
  \mathcal{L}_{\Xi \Xi \rho} &=& g_{\rho 2}
  \,{\Xi}^{\dagger} \vec{\tau} \cdot \vec{\rho}^{0}  {\Xi} \nonumber \\
  &-& \frac{f_{\rho 2}}{4 M} \,\epsilon_{ijk} {\Xi}^{\dagger} \sigma_k
  \vec{\tau} \cdot \left( \partial_i \vec{\rho}_j -
  \partial_j \vec{\rho}_i
  \right)  {\Xi} \, , \\
  \mathcal{L}_{\Xi \Xi \omega} &=& g_{\omega 2}
  \,{\Xi}^{\dagger} {\omega}^{0} {\Xi} \nonumber \\
  &-& \frac{f_{\omega 2}}{4 M} \,\epsilon_{ijk}\,{\Xi}^{\dagger} \sigma_{k}
  \,\left( \partial_i {\omega}_j - \partial_j {\omega}_i
  \right) {\Xi} \, .
\end{eqnarray}

\subsection{The OBE Potential}

The general form of the $P\, \Xi$ and $P^* \, \Xi$ OBE potential is
\begin{eqnarray}
  V = \zeta \, V_{\pi} + V_{\sigma} + V_{\rho} + \zeta \, V_{\omega} \, ,
\end{eqnarray}
where the subscript indicates from which meson comes the contribution
($\pi$, $\sigma$, $\rho$ or $\omega$) and with $\zeta = \pm 1$ a sign.
We take the convention that
\begin{eqnarray}
  \zeta = +1 && \mbox{for $P \Xi$ and $P^* \Xi$} \, , \\
  \zeta = -1 && \mbox{for $\bar{P} \Xi$ and $\bar{P}^* \Xi$} \, .
\end{eqnarray}
If the vector meson and the hadrons are point-like, their vertices
can be directly computed from the Lagrangian and we end up with
the following potentials in momentum space
\begin{eqnarray}
  V_{\pi}(\vec{q})
  &=& \vec{\tau}_1 \cdot \vec{\tau}_2\,\frac{g_1 g_2}{2 f_{\pi}^2}\,
  \frac{a_1 \cdot \vec{q}\, \vec{\sigma}_2 \cdot \vec{q}}{q^2 + m_{\pi}^2}
  \, , \\
  V_{\sigma}(\vec{q}) &=& -\frac{g_{\sigma 1} g_{\sigma 2}}{q^2 + m_{\sigma}^2} \, \\
  V_{\rho}(\vec{q}) &=& \vec{\tau}_1 \cdot \vec{\tau}_2\, \Big[
    \frac{g_{\rho 1} g_{\rho 2}}{q^2 + m_{\rho}^2} \nonumber \\
    &-&
    \frac{f_{\rho 1}}{2 M_1}\,\frac{f_{\rho 2}}{2 M_2}\,
    \frac{(a_1 \times \vec{q}) \cdot
      (\vec{\sigma}_2 \times \vec{q})}{q^2 + m_{\rho}^2} \Big]
   \, , \\
   V_{\omega}(\vec{q}) &=& -\frac{g_{\omega 1} g_{\omega 2}}{q^2 + m_{\omega}^2}
   \nonumber \\
    &+&
   \frac{f_{\omega 1}}{2 M_1}\,\frac{f_{\omega 2}}{2 M_2}\,
   \frac{(a_1 \times \vec{q}) \cdot
      (\vec{\sigma}_2 \times \vec{q})}{q^2 + m_{\omega}^2}
  \, .
\end{eqnarray}
In the expressions the subscript $1$ and $2$ is used
for the $P$/$P^*$ heavy meson and $\Xi$ baryon respectively.
For  $\vec{a}_1$ we take
\begin{eqnarray}
  \vec{a}_1 = 0\,\, && \mbox{for $P$} , \\
  \vec{a}_1 = \vec{S}_1 && \mbox{for $P^*$} ,
\end{eqnarray}
with $\vec{S}$ the spin-1 angular momentum matrices.
For the pion decay constant we take the $f_{\pi} = 132\,{\rm MeV}$ normalization.
The choice of sign for the momentum space potential is such that 
the Lippmann-Schwinger equation reads $T = V + V G_0 T$,
where the T-matrix is in turn normalized so that for 
zero-energy scattering $T \to 2 \pi \, a_0 / \mu$,
with $a_0$ the scattering length and
$\mu$ the reduced mass of the system.
That is, we are using the standard non-relativistic normalization
which is also used in the two-nucleon system,
see for instance Ref.~\cite{Machleidt:1989tm}.

We can take into account the finite size of hadrons
by including form factors in the calculation, that is
\begin{eqnarray}
  V_M(\vec{q}, \Lambda) = V_M(\vec{q})\,
  F_1(q, m, \Lambda_1)\,F_2(q, m, \Lambda_1)
  \, ,
\end{eqnarray}  
where $F_1$ and $F_2$ are the form factors corresponding to vertex $1$ and $2$,
i.e. the $P$/$P^*$ heavy meson and the $\Xi$ baryon.
The form factor can depend on the momentum transfer $q$, the mass of
the exchanged meson $m$ and a cut-off $\Lambda$.
Here we will use a monopolar form factor of the type
\begin{eqnarray}
  F(q, m, \Lambda)=\frac{\Lambda^{2}-m^2}{\Lambda^{2}-{q}^2} \label{Eq:FF} \, ,
\end{eqnarray}
for both vertices, where $q^2 = q_0^2 - \vec{q}\,^2$ is the 4-momentum of the
exchanges meson.

In configuration space and for point-like interactions
the components of the OBE potential take the form
\begin{eqnarray}
  V_{\pi}(\vec{r}) &=&
  -\vec{\tau}_1 \cdot \vec{\tau}_2\,\frac{g_1 g_2}{6 f_{\pi}^2}\,\Big[
    - \vec{a}_1 \cdot \vec{\sigma}_2\,\delta(\vec{r})
    \nonumber \\ && \quad
    + \vec{a}_1 \cdot \vec{\sigma}_2\,m_{\pi}^3\,W_Y(m_{\pi} r)
    \nonumber \\ && \quad
    + S_{12}(\vec{r})\,m_{\pi}^3\,W_T(m_{\pi} r) \Big] \, , \\
  V_{\sigma}(\vec{r}) &=& -{g_{\sigma 1} g_{\sigma 2}}\,m_{\sigma}\,W_Y(m_{\sigma} r)
  \, , \\
  V_{\rho}(\vec{r}) &=& \vec{\tau}_1 \cdot \vec{\tau}_2\,\Big[
    {g_{\rho 1} g_{\rho_2}}\,m_{\rho}\,W_Y(m_{\rho} r) \nonumber \\
    && \quad + \frac{f_{\rho 1}}{2 M_1}\,\frac{f_{\rho 2}}{2 M_2}\,\Big(
    -\frac{2}{3}\,\vec{a}_1 \cdot \vec{\sigma}_2 \, \delta(\vec{r})
    \nonumber \\ && \qquad
    +\frac{2}{3}\,\vec{a}_1 \cdot \vec{\sigma}_2 \, m_{\rho}^3 \, W_Y(m_{\rho} r)
    \nonumber \\ && \qquad
    -\frac{1}{3}\,S_{12}(\hat{r})\, m_{\rho}^3 \, W_T(m_{\rho} r) \,\,
    \Big) \, \Big] \, , \\
    V_{\omega}(\vec{r}) &=& 
    -{g_{\omega 1} g_{\omega_2}}\,m_{\omega}\,W_Y(m_{\omega} r) \nonumber \\
    && \quad - \frac{f_{\omega 1}}{2 M_1}\,\frac{f_{\omega 2}}{2 M_2}\,\Big(
    -\frac{2}{3}\,\vec{a}_1 \cdot \vec{\sigma}_2 \, \delta(\vec{r})
    \nonumber \\ && \qquad
    +\frac{2}{3}\,\vec{a}_1 \cdot \vec{\sigma}_2 \,
    m_{\omega}^3 \, W_Y(m_{\omega} r)
    \nonumber \\ && \qquad
    -\frac{1}{3}\,S_{12}(\hat{r})\, m_{\omega}^3 \, W_T(m_{\omega} r) \,\,
    \Big) \, , 
\end{eqnarray}
where the functions $W_Y(x)$ and $W_T(x)$ are defined as
\begin{eqnarray}
  W_Y(x) &=& \frac{e^{-x}}{4\pi x} \, , \\
  W_T(x) &=& \left( 1 + \frac{3}{x} + \frac{3}{x^2} \right)
  \,\frac{e^{-x}}{4\pi x} \, .
\end{eqnarray}
The effects of the finite size of the hadrons is easy to take into account
by making the changes
\begin{eqnarray}
  \delta(r) &\to& m^3\,d(x,\lambda) \, , \\
  W_Y(x) &\to& W_Y(x, \lambda) \, , \\
  W_T(x) &\to& W_T(x, \lambda) \, , 
\end{eqnarray}
where $\lambda = \Lambda / m$.
For the monopolar form factor of Eq.~(\ref{Eq:FF}) we end up with
\begin{eqnarray}
  d(x, \lambda) &=& \frac{(\lambda^2 - 1)^2}{2 \lambda}\,
  \frac{e^{-\lambda x}}{4 \pi} \, , \\
  W_Y(x, \lambda) &=& W_Y(x) - \lambda W_Y(\lambda x) \nonumber \\ && -
  \frac{(\lambda^2 - 1)}{2 \lambda}\,\frac{e^{-\lambda x}}{4 \pi} \, , \\
  W_T(x, \lambda) &=& W_T(x) - \lambda^3 W_T(\lambda x) \nonumber \\ && -
  \frac{(\lambda^2 - 1)}{2 \lambda}\,\lambda^2\,
  \left(1 + \frac{1}{\lambda x} \right)\,\frac{e^{-\lambda x}}{4 \pi} \, .
\end{eqnarray}

As a matter of fact the structure of the OBE potential presented
here is exceedingly simple.
We can write it as a sum of a central, spin-spin and tensor component
\begin{eqnarray}
  V(\vec{r}) &=&
  V_C(r) + \vec{a}_1 \cdot \vec{\sigma}_2 \, V_S(r) + S_{12}(\hat{r})\,V_T(r)
  \, ,
\end{eqnarray}
where for point-like interactions we have
\begin{eqnarray}
  V_C(r) &=& -{g_{\sigma 1} g_{\sigma 2}}\,m_{\sigma}\,W_Y(m_{\sigma} r)
  \nonumber \\ && 
  + \vec{\tau}_1 \cdot \vec{\tau}_2\,
  {g_{\rho 1} g_{\rho_2}}\,m_{\rho}\,W_Y(m_{\rho} r)
  \nonumber \\ && 
  - \zeta \, {g_{\omega 1} g_{\omega_2}}\,m_{\omega}\,W_Y(m_{\omega} r) \, , \\
  V_S(r) &=&
  - \zeta\,\vec{\tau}_1 \cdot \vec{\tau}_2\,\frac{g_1 g_2}{6 f_{\pi}^2}\,
  \left[ -\delta(\vec{r}) + m_{\pi}^3\,W_Y(m_{\pi} r) \right]
  \nonumber \\ && +
  \frac{2}{3}\,
  \vec{\tau}_1 \cdot \vec{\tau}_2\,
  \frac{f_{\rho 1}}{2 M_1}\,\frac{f_{\rho 2}}{2 M_2}\,
  \left[ -\delta(\vec{r}) + m_{\rho}^3\,W_Y(m_{\rho} r) \right]
  \nonumber \\ && -
  \frac{2}{3}\,\zeta\,\frac{f_{\omega 1}}{2 M_1}\,\frac{f_{\omega 2}}{2 M_2}\,
  \left[ -\delta(\vec{r}) + m_{\omega}^3\,W_Y(m_{\omega} r) \right] \, , \\
  V_T(r) &=&
  -\zeta\,\vec{\tau}_1 \cdot \vec{\tau}_2\,\frac{g_1 g_2}{6 f_{\pi}^2}\,
  m_{\pi}^3\,W_T(m_{\pi} r)
  \nonumber \\ && -
  \frac{1}{3}\,\vec{\tau}_1 \cdot \vec{\tau}_2\,
  \frac{f_{\rho 1}}{2 M_1}\,\frac{f_{\rho 2}}{2 M_2}\,m_{\rho}^3\,W_T(m_{\rho} r)
  \nonumber \\ && +
  \frac{1}{3}\,\zeta\,\frac{f_{\omega 1}}{2 M_1}\,\frac{f_{\omega 2}}{2 M_2}\,
  m_{\omega}^3\,W_T(m_{\omega} r) \, ,
\end{eqnarray}
while for finite-size hadrons we substitute $\delta(r)$, $W_Y(x)$ and $W_T(x)$
by their finite-size versions.

\subsection{Couplings}

For the $D$ and $D^*$ heavy mesons the axial coupling with the pion we take
\begin{eqnarray}
  g_1 = 0.60 \, ,
\end{eqnarray}
which is compatible with $g_1 = 0.59 \pm 0.01 \pm 0.07$ as
extracted from the $D^* \to D \pi$ decay~\cite{Ahmed:2001xc,Anastassov:2001cw}.

The coupling to the $\sigma$ meson, in the case of the nucleon-nucleon
interaction, can be determined from
the non-linear sigma model~\cite{GellMann:1960np} yielding
\begin{eqnarray}
  g_{\sigma NN} = \sqrt{2}\,\frac{M_N}{f_{\pi}} \simeq 10.2 \, .
\end{eqnarray}
For the case of the $D$, $D^*$ mesons and $\Xi$ baryons we can estimate
the coupling to the $\sigma$ from the quark model. If we assume
that $\sigma$ only couples to the $u$ and $d$ quarks, we expect
\begin{eqnarray}
  g_{\sigma 1} = g_{\sigma 2} = \frac{g_{\sigma NN}}{3} \simeq 3.4 \, .
\end{eqnarray}

We can also deduce from SU(3) flavour symmetry and the OZI rule that
\begin{eqnarray}
  g_{\rho 1} &=& g_{\omega 1} \, , \\
  g_{\rho 2} &=& g_{\omega 2} \, . 
\end{eqnarray}
From the universality of the $\rho$ couplings
(Sakurai's universality~\cite{Sakurai:1960ju})
and the KSFR (Kawarabayashi-Suzuki-Fayyazuddin-Riazuddin)
relation~\cite{Kawarabayashi:1966kd,Riazuddin:1966sw} we expect
\begin{eqnarray}
  g_{\rho 1} = g_{\rho 2} = \frac{m_{\rho}}{2 f_{\pi}^2} \simeq 2.9 \, .
\end{eqnarray}
Yet there might be deviations from this value.
Regarding the $\rho$ coupling to the heavy mesons,
Casalbuoni et al.~\cite{Casalbuoni:1992dx} indicate that 
\begin{eqnarray}
  g_{\rho 1} = \beta \frac{m_{\rho}}{2 f_{\pi}^2} \simeq 2.6 \, ,
\end{eqnarray}
where $\beta = 0.9$. The $\rho$ coupling thus obtained actually coincides
with lattice QCD calculations in the heavy quark limit~\cite{Detmold:2012ge},
which yield $g_{\rho 1} = 2.6 \pm 0.1 \pm 0.4$.
For the $\rho$ and $\omega$ coupling to the cascade,
there is also the possibility of obtaining it from the nucleon-nucleon case.
In that case the relevant relations are (see the Appendix)
\begin{eqnarray}
  g_{\rho 2} &=& g_{\rho NN} \, , \\
  g_{\omega 2} &=& \frac{1}{3}\,g_{\omega NN} \, .
\end{eqnarray}
In the nucleon-nucleon case the SU(3) + OZI relation is
\begin{eqnarray}
  g_{\omega NN} = 3\,g_{\rho NN} \, ,
\end{eqnarray}
which is compatible with the analogous relation for the cascade baryon
once we take into account the quark model.
Yet nuclear potentials usually violate the previous relation,
requiring a $g_{\omega NN} \sim 20$ or larger,
a discrepancy which has been long known in OBE models and
usually attributed to the fact that the $g_{\omega NN}$ used
in nuclear potentials might indeed also account for some of
the short-range quark-gluon exchanges~\cite{Machleidt:1989tm}.
Yet, this discrepancy can be understood in more modern terms within
the renormalization ideas that have become commonplace after
the advent of chiral EFT.
The explanation lies in the fine-tuning nature of
the nucleon-nucleon interaction, which translates
into a fine-tuning of the $\omega$ coupling.
In fact the $\omega$ coupling provides a very important central contribution
to the nuclear force, which is responsible in a large part for the concrete
values of the scattering lengths of the singlet and triplet channels.
By combining the OBE model with modern renormalization techniques
this discrepancy disappears and the SU(3) relation
is recovered~\cite{Cordon:2009pj}.
These findings indicate that the use of the SU(3) relations is the most
judicious choice to determine the $g_{\omega}$ couplings, at least
for exploratory studies of prospective hadron molecules where
we are not trying to fit fine-tuned systems.

For $f_{\rho}$ and $f_{\omega}$ the estimations in the case of the charmed mesons
are as follows. SU(3) and the OZI rule imply that
\begin{eqnarray}
  f_{\rho 1} = f_{\omega 1} \, .
\end{eqnarray}
This relation appears automatically if the lagrangians are written in terms
of the vector meson nonet.
Meanwhile vector meson dominance applied to the weak decays of
the charmed mesons~\cite{Casalbuoni:1992dx} bring us to
\begin{eqnarray}
  \frac{f_{\rho 1}}{2 M_1} &=&
  {2 \lambda}\,\frac{m_{\rho}}{2 f_{\pi}}
   \\
  &=& \kappa_{\rho 1}\,\frac{g_{\rho 1}}{2 m_H}\, ,
\end{eqnarray}
with $|\lambda| = 0.60 \pm 0.11\,{\rm GeV}^{-1}$ and where in the second line
we have written the coupling of the $\rho$ in the normalization
for which we take $M_1 = m_H$ with $m_H$ the mass of
the charmed meson.
If we take $m_H = m_D$, $g_{\rho 1} = 2.6$ and assume that $\lambda$ is positive
we obtain
\begin{eqnarray}
  \kappa_{\rho 1} = 4.5 \pm 0.8 \, .
\end{eqnarray}

For the cascade the estimations are more involved.
The reason is that the coupling of the $\rho$ meson to the octet baryons
depends in general on two coupling constants, the symmetric and
antisymmetric octet couplings~\footnote{
  The electric-type coupling of the $\rho$ meson --- the $g_{\rho}$ coupling ---
  is an exception because of its universal character.}.
In the case of the nucleon-nucleon interaction it is possible to use
single vector meson dominance to obtain the relation
\begin{eqnarray}
  f_{\rho NN} &=& \kappa_{\rho NN}\,g_{\rho NN}  \, ,\\
  f_{\omega NN} &=& \kappa_{\omega NN}\,g_{\omega NN} \, ,
\end{eqnarray}
with $\kappa_{\rho NN} = \mu_p - \mu_n - 1$,
$\kappa_{\omega NN} = \mu_p + \mu_n - 1 $,
yielding $\kappa_{\rho NN} = 3.7$ and $\kappa_{\omega NN} = -0.1$.
This idea can be adapted to the $D$ and $D^*$ charmed mesons and the $\Xi$
baryon, in which case we have
$\kappa_{\rho 2} = \mu_{\Xi^0} - \mu_{\Xi^{-}} - 1$ and
$\kappa_{\omega 2} = \mu_{\Xi^0} + \mu_{\Xi^{-}} + 1$.
The application of the previous idea implicitly assumes the convention
$M_2 = m_{\Xi}$, which we will follow onwards.
From the experimental values $\mu_{\Xi^0} = -0.6507(25)$ and
$\mu_{\Xi^-} = -1.250(14)$ listed in the PDG~\cite{Patrignani:2016xqp}
we obtain $\kappa_{\rho 2} = -0.401$ and $\kappa_{\omega 2} = -0.901$.
Other possibility is to contrain them from the quark model,
in which case we obtain
\begin{eqnarray}
  1 + \kappa_{\rho 2} &=& -\frac{1}{3}\,\frac{m_{\Xi}}{m_N}\,
  \left( 1 + \kappa_{\rho NN} \right) \, , \\
  1 + \kappa_{\omega 2} &=& -\frac{1}{5}\,\frac{m_{\Xi}}{m_N}\,
  \left( 1 + \kappa_{\rho NN} \right) \, ,
\end{eqnarray}
which yield $\kappa_{\rho 2} = -3.2$ and $\kappa_{\omega 2} = -1.3$,
in stark contrast with the previous estimations.
We can also consider the family of phenomenological soft-core potentials
by the Nijmegen group~\cite{Rijken:1998yy,Rijken:2006ep,Rijken:2006kg,Nagels:2014qqa,Nagels:2015lfa,Nagels:2015dia}
which also contain estimations for the electric and magnetic couplings
of the vector mesons with the cascade.
In this case we have $\kappa_{\rho 2} = -2.0, -0.7, -0.3$
$\kappa_{\omega 2} = -1.1, -1.9, -2.3$
for the ESC04a, ESC04d and ESC08c potentials respectively~\footnote{
  We have simply divided the magnetic and electric couplings
  $\kappa = f/g$ for these potentials.},
though it is worth noticing that $\kappa_{\omega 2}$ is obtained
from a value of the omega coupling $g_{\omega 2} \sim (2-3)\,g_{\rho 2}$
that is considerably larger than the SU(3) + OZI rule expectation.
From the previous discussion it is apparent that there is a considerable
level of uncertainty in $\kappa_{\rho 2}$ and $\kappa_{\omega 2}$.
For simplicity we will use the values
\begin{eqnarray}
\kappa_{\rho 2} = \kappa_{\omega 2} = -1.5 \, .
\end{eqnarray}
This choice is similar to the geometric mean of the previous determinations
($\kappa_{\rho 2} = -1.3$ and $\kappa_{\omega 2} = -1.5$)
and to the values we obtain when we compute the cascade magnetic moments
at tree level in chiral perturbation theory
($\mu_{\Xi^0} = -1.60$ and $\mu_{\Xi^-} = -0.97$~\cite{Geng:2008mf}
yielding $\kappa_{\rho 2} = -1.63$ and $\kappa_{\omega 2} = -1.57$).
We review the set of parameters we use in the OBE potential
in Tables \ref{tab:mass} and \ref{tab:couplings}.

\begin{table}[!h]
\begin{tabular}{|ccc|}
\hline\hline
  Hadron  & $I\,(J^{P})$  & M (MeV) \\
  \hline
  $N$ & $\frac{1}{2}$ $({\frac{1}{2}}^{+})$ & 938 \\
    $\Xi$ & $\frac{1}{2}$ $({\frac{1}{2}}^{+})$ & 1318 \\ \hline
  $D$ & $\frac{1}{2}$ $({0}^{-})$ & 1867 \\
  $D^*$ & $\frac{1}{2}$ $({1}^{-})$ & 2009 \\
  $B$ & $\frac{1}{2}$ $({0}^{-})$ & 5279 \\
  $B^*$ & $\frac{1}{2}$ $({1}^{-})$ & 5325 \\ \hline
  $\Xi_{cc}$ & $\frac{1}{2}({\frac{1}{2}}^{+})$ & 3621 \\
  $\Xi_{bb}$ & $\frac{1}{2}({\frac{1}{2}}^{+})$ & 10127 \\
  \hline\hline 
\end{tabular}
\caption{Masses and quantum numbers of the hadrons
  from which we form molecules in the present work
  (plus the nucleon). For the $N$ and heavy mesons we use the isospin average
  of the masses listed in the PDG~\cite{Patrignani:2016xqp}.
  For the $\Xi_{cc}$ we use the experimental value of the $\Xi_{cc}^{++}$ mass
  from the LHCb collaboration~\cite{Aaij:2017ueg},
  and for the $\Xi_{bb}$ we use the lattice QCD determination of
  Ref.~\cite{Lewis:2008fu}. We round the numbers
  at the ${\rm MeV}$ level.}
\label{tab:mass}
\end{table}

\begin{table}[!h]
\begin{tabular}{|ccc|}
  \hline\hline
  Meson  & $I^{G}\,(J^{PC})$  & M (MeV) \\
  \hline
  $\pi$ & $1^{-}$ $({0}^{-+})$ & 138 \\
  $\sigma$ & $0^{+}$ $({0}^{++})$ & 600 \\
  $\rho$ & $1^{+}$ $({1}^{--})$ & 770 \\
  $\omega$ & $0^{-}$ $({1}^{--})$ & 780 \\
  \hline \hline \hline
  Coupling  & $D$/$D^*$  & $\Xi$ \\
  \hline
  $g$ & 0.60 & -0.25 \\
  $g_{\sigma}$ & 3.4 & 3.4 \\
  $g_{\rho}$ & 2.6 & 2.9\\
  $g_{\omega}$ & 2.6 & 2.9 \\
  $\kappa_{\rho}$ & 4.5 &  -1.5\\
  $\kappa_{\omega}$ & 4.5 & -1.5\\
  $M$ & 1867 & 1318 \\
  \hline \hline
\end{tabular}
\caption{Masses, quantum numbers and couplings of the light mesons
  of the OBE model ($\pi$, $\sigma$, $\rho$, $\omega$).
  For the magnetic-type coupling of the $\rho$ and $\omega$ vector mesons
  we have used the decomposition
  $f_{\rho (\omega)} = \kappa_{\rho (\omega)}\,g_{\rho (\omega)}$.
  $M$ refers to the mass scale involved in the magnetic-type couplings.
}
\label{tab:couplings}
\end{table}

\subsection{$D\Xi$ and $D^*\Xi$ Wave Function}

For the molecules we are considering here, the total wave function is
the product of the isospin and spin-spatial wave functions
\begin{eqnarray}
  | \Psi \, \rangle = | I M_I \rangle \, |\psi_{J M}(\vec{r}) \, \rangle \, .
\end{eqnarray}
where $J$ refers to the total angular momentum of
the molecule under consideration.
The isospin wave function comes from the coupling of the isospin of
the two hadrons in the molecule
\begin{eqnarray}
  | I M_I \rangle = \sum_{M_{I1} M_{I2}} \langle I_1 M_{I1} I_2 M_{I2} | I M \rangle
  \, | I_1 M_{I1} \rangle \, | I_2 M_{I2} \rangle \, .
\end{eqnarray}
The only subtlely in the isospin wave function is when the hadron contains
a light anti-quark $\bar q$, for which there will be a minus sign for one of
the components of the isospin multiplet.
For instance, if we consider the $\bar{D}$ (${\bar c} q$)
and the $D$ ($c \bar q$) we have
\begin{eqnarray}
  \bar{D} =
  \begin{pmatrix}
    \bar{D}^0 \\
    D^{-}
  \end{pmatrix}
  \quad \mbox{and} \quad
  {D} =
  \begin{pmatrix}
    D^{+} \\
    -\bar{D}^0
  \end{pmatrix} \, ,
\end{eqnarray}
where the upper and lower components represent
the $| \frac{1}{2} +\frac{1}{2} \rangle$ and
$| \frac{1}{2} -\frac{1}{2} \rangle$
isospinors respectively.
In constrast, for the cascade we simply have
\begin{eqnarray}
  \Xi =
  \begin{pmatrix}
    \Xi^0 \\
    \Xi^{-}
  \end{pmatrix} \, ,
\end{eqnarray}
where we are using the same prescription for the isospinors.

For the part of the wavefunction containing the spatial and spin pieces,
we can express it as a sum of different components
with the same parity and total angular momentum,
i.e. a partial wave expansion
\begin{eqnarray}
  |\psi_{J M}(\vec{r}) \, \rangle = \sum_{L S} \psi_{L S J}(r)\,
  | {}^{2S+1}L_J \rangle \, ,
\end{eqnarray}
where the sum over angular momentum only comprises even or odd $L$ depending
on the parity of the molecule $P = (-1)^L$.
For the partial wave projection we have adopted
the spectroscopic notation $^{2S+1}L_J$, where $S$ is the total spin,
$L$ the orbital angular momentum and $J$ the total angular momentum.
We define the $| {}^{2S+1}L_J \rangle$ as follows
\begin{eqnarray}
  |{}^{2S+1}L_{J}\rangle &=& \sum_{M_{S},M_L}
  \langle L M_L S M_S | J M \rangle \, | S M_S \rangle \, Y_{L M_{L}}(\hat{r})
  \, ,
\end{eqnarray}
where $\langle L M_L S M_S | J M \rangle$ is a Clebsch-Gordan coefficient,
$Y_{L M_L}(\hat{r})$ a spherical harmonic and
$| S M_S \rangle$ the spin wavefunction,
which in turn can be defined as
\begin{eqnarray}
  | S M_S\rangle &=& \sum_{M_{S1},M_{S2}}
  \langle S_1 M_{S1} S_2 M_{S2} | S M_S \rangle \,
  | S_1 M_{S1} \rangle \, | S_2 M_{S2} \rangle \, , \nonumber \\
\end{eqnarray}
with $| S_1 M_{S1} \rangle$, $| S_2 M_{S2} \rangle$ are the spin wavefunction
of particle $1$ and $2$.

The mixing of partial waves with the same $J$ but different $S$/$L$
requires the tensor force.
The coupling only happens in the $D^* \Xi$ and ${\bar D}^* \Xi$ cases,
because for $D \Xi$ and ${\bar D} \Xi$ the tensor force disappears.
As molecular states are expected to be more probable for S-waves,
we will consider only the following partial waves:
\begin{eqnarray}
  | D \, \Xi (J=\frac12) \rangle &=&
  |{}^{2} {S}_{\frac{1}{2}} \rangle \, , \label{eq:pw1} \\
  | D^* \, \Xi (J=\frac12) \rangle &=&
  \left \{
  |{}^{2} {S}_{\frac{1}{2}} \rangle ,
  |{}^{4} {D}_{\frac{1}{2}} \rangle
  \right \}
  \, , \label{eq:pw2} \\
  | D^* \, \Xi (J=\frac32)\rangle &=&
  \left \{
  |{}^{4} {S}_{\frac{3}{2}} \rangle,
  |{}^{2} {D}_{\frac{3}{2}} \rangle,
  |{}^{4} {D}_{\frac{3}{2}} \rangle \right \}
  \, . \label{eq:pw3}
\end{eqnarray}
The evaluation of the spin-spin and tensor operators for these channels
can be found in Table \ref{tab:tensor}.

\begin{table}[t]
\begin{tabular}{|c|cc|}
\hline\hline
&  $\vec{a}_1 \cdot \vec{\sigma}_2 $ & $S_{12}$  \\ \hline
$D \Xi (J=\frac{1}{2})$ & 0 & 0 \\ \hline
$D^* \Xi (J=\frac{1}{2})$ &
$\left(\begin{matrix}
-2 & 0 \\
0 & 1%
\end{matrix}\right)$  & $\left(\begin{matrix}
0 & -\sqrt{2} \\
-\sqrt{2} &-2%
\end{matrix}\right)$\\  \hline
$D^* \Xi (J=\frac{3}{2})$ &
$\left(\begin{matrix}
1 & 0&0 \\
0 & -2&0 \\
0& 0& 1%
\end{matrix}\right)$& $\left(\begin{matrix}
0 & 1& 2 \\
1 &0 &-1\\
2 &-1 &0
\end{matrix}\right)$  \\
\hline\hline
\end{tabular}
\centering \caption{Matrix elements of the spin-spin and
  tensor operator for the partial waves we are considering
  in this work, see Eqs.~(\ref{eq:pw1}), (\ref{eq:pw2}) and (\ref{eq:pw3})
  for the definitions.} \label{tab:tensor}
\end{table}

\subsection{The Extension to $\Xi_{cc}$ and $\Xi_{bb}$ Baryons}

We can extend the OBE model to the doubly heavy baryons in two ways.
The first is the quark model, in which case we derive the interactions
of the $\Xi_{cc}$ and $\Xi_{bb}$ with the light mesons
from the ones for the ${\Xi}$.
The second is heavy antiquark-diquark symmetry~\cite{Savage:1990di,Hu:2005gf,Guo:2013xga},
in which case the derivation is from the heavy meson interactions.

In the quark model we expect the strange quark to act
as an expectator in what refers to the couplings
to the $\pi$, $\sigma$, $\rho$ and $\omega$
light mesons.
From this the OBE lagrangian for the $\Xi_{cc}$ and $\Xi_{bb}$ baryons is
\begin{eqnarray}
    \mathcal{L}_{\Xi_{QQ} \Xi_{QQ} \pi} &=&
    \frac{g_2}{\sqrt{2} f_{\pi}}\,
       {\Xi_{QQ}}^{\dagger}
       \vec{\sigma} \cdot \nabla ( \vec{\tau} \cdot \vec{\pi})  
       \Xi_{QQ} \, , \\
  \mathcal{L}_{\Xi_{QQ} \Xi_{QQ} \sigma} &=& g_{\sigma 2}
  \,\Xi_{QQ}^{\dagger} \sigma \Xi_{QQ} \, , \\
  \mathcal{L}_{\Xi_{QQ} \Xi_{QQ} \rho} &=& g_{\rho 2}
  \,{\Xi}_{QQ}^{\dagger} \vec{\tau} \cdot \vec{\rho}^{0}  {\Xi_{QQ}} \nonumber \\
  &-& \frac{f_{\rho 2}}{4 M_2} \,\epsilon_{ijk}
    {\Xi_{QQ}}^{\dagger} \sigma_k
    \vec{\tau} \cdot \left( \partial_i \vec{\rho}_j -
    \partial_j \vec{\rho}_i
    \right)  {\Xi_{QQ}} \, , \nonumber \\ \\
  \mathcal{L}_{\Xi_{QQ} \Xi_{QQ} \omega} &=& g_{\omega 2}
  \,{\Xi}^{\dagger} {\omega}^{0} {\Xi} \nonumber \\
  &-& 
  \frac{f_{\omega 2}}{4 M_2} \,\epsilon_{ijk}\,{\Xi_{QQ}}^{\dagger} \sigma_{k}
  \,\left( \partial_i {\omega}_j - \partial_j {\omega}_i
  \right) {\Xi_{QQ}} \, , \nonumber \\
\end{eqnarray}
where $M_2$ is the same as in the original lagrangian for the cascade.
In short, the couplings are the same as for the $\Xi$.

Heavy heavy antiquark-diquark symmetry (HADS) is a manifestation of
heavy quark symmetry which states that a heavy quark pair behaves
as a heavy antiquark~\cite{Savage:1990di}.
According to this symmetry the couplings of the $\Xi_{cc}$ and $\Xi_{bb}$
baryons can be deduced from those of the ${\bar D}$, ${\bar D}^*$ and
${\bar B}$, ${\bar B}^*$ heavy antimesons~\cite{Hu:2005gf,Guo:2013xga}.
The application of HADS can actually be encapsulated
in the following two relations between the lagrangian
for the $\bar{P}$ and $\bar{P}^*$ heavy antimesons and
the $\Xi_{QQ}$ doubly heavy baryons
\begin{eqnarray}
  {\rm Tr}\left[ {\bar H}^{\dagger} {\bar H}\right] &\to&
  {\Xi_{QQ}}^{\dagger} \, {\Xi_{QQ}} \, , \\
  {\rm Tr}\left[ {\bar H}^{\dagger} \vec{\sigma} {\bar H}\right] &\to&
  - \frac{1}{3}\,{\Xi_{QQ}}^{\dagger} \, \vec{\sigma} \, {\Xi_{QQ}} \, ,
\end{eqnarray}
where the bar over the $H$ field indicates
that we are dealing the heavy antimeson superfield.
From this, the OBE lagrangian for the heavy mesons and changing the sign
of the $\pi$ and $\omega$ contributions to take into account
their G-parity, we arrive at
\begin{eqnarray}
  \mathcal{L}_{\Xi_{QQ} \Xi_{QQ} \pi} &=&
  -\frac{1}{3}\,\frac{g_1}{\sqrt{2} f_{\pi}}\,
       {\Xi_{QQ}}^{\dagger}
       \vec{\sigma} \cdot \nabla ( \vec{\tau} \cdot \vec{\pi})  
       \Xi_{QQ} \, , \\
  \mathcal{L}_{\Xi_{QQ} \Xi_{QQ} \sigma} &=& g_{\sigma 1}
  \,\Xi_{QQ}^{\dagger} \sigma \Xi_{QQ} \, , \\
  \mathcal{L}_{\Xi_{QQ} \Xi_{QQ} \rho} &=& g_{\rho 1}
  \,{\Xi}_{QQ}^{\dagger} \vec{\tau} \cdot \vec{\rho}^{0}  {\Xi_{QQ}} \nonumber \\
  &+& \frac{1}{3}\,\frac{f_{\rho 1}}{4 M_1} \,\epsilon_{ijk}
    {\Xi_{QQ}}^{\dagger} \sigma_k
    \vec{\tau} \cdot \left( \partial_i \vec{\rho}_j -
    \partial_j \vec{\rho}_i
    \right)  {\Xi_{QQ}} \, , \nonumber \\ \\
  \mathcal{L}_{\Xi_{QQ} \Xi_{QQ} \omega} &=& g_{\omega 1}
  \,{\Xi}^{\dagger} {\omega}^{0} {\Xi} \nonumber \\
  &+& \frac{1}{3}\,
  \frac{f_{\omega 1}}{4 M_1} \,\epsilon_{ijk}\,{\Xi_{QQ}}^{\dagger} \sigma_{k}
  \,\left( \partial_i {\omega}_j - \partial_j {\omega}_i
  \right) {\Xi_{QQ}} \, . \nonumber \\
\end{eqnarray}

The comparison with the lagrangian derived from the quark model
entails the following HADS predictions
\begin{eqnarray}
  {\left( g_2 \right)}_{\rm HADS} &=& -\frac{g_1}{3} \, , \\
  {\left( g_{\sigma 2} \right)}_{\rm HADS} &=& g_{\sigma 1} \, , \\
  {\left( g_{\rho 2} \right)}_{\rm HADS} = g_{\rho 1} \, & , & \,
  {\left( \frac{f_{\rho 2}}{2 M_2} \right)}_{\rm HADS}
  = -\frac{1}{3}\,\frac{f_{\rho 1}}{2 M_1} \, , \\
  {\left( g_{\omega 2} \right)}_{\rm HADS} = g_{\omega 1} \, & , & \,
  {\left( \frac{f_{\omega 2}}{2 M_2} \right)}_{\rm HADS}
  = -\frac{1}{3}\,\frac{f_{\omega 1}}{2 M_1}
\end{eqnarray}
which can actually be checked against the quark model expectations.
Owing to the choices of the couplings made before, we only have to
compare five couplings: $g_2$, $g_{\rho 2}$, $g_{\omega 2}$,
$f_{\rho 2}$ and $f_{\omega 2}$. These comparisons are reduced
to three as $g_{\rho 2} = g_{\omega 2}$ from SU(3) + OZI and
$f_{\rho 2} = f_{\omega 2}$ because of the choice we have made.
For the axial coupling we have
\begin{eqnarray}
  {\left( g_2 \right)}_{\rm QM} = -0.25 \, &\mbox{vs}& 
  {\left( g_2 \right)}_{\rm HADS} = -0.20 \, ,
\end{eqnarray}
which are actually very similar.
For the $\rho$ electric-type couplings we have
\begin{eqnarray}
  {\left( g_{\rho 2} \right)}_{\rm QM} = 2.9 \, &\mbox{vs}& 
  {\left( g_{\rho 2} \right)}_{\rm HADS} = 2.6 \, , 
\end{eqnarray}
which differ by a $10\,\%$ only.
For the magenetic-type couplings,
if we employ $M_2 = m_{\Xi}$ in the doubly heavy sector, the comparison
can be directly made in terms of $\kappa_{\rho 2}$ and $\kappa_{\omega 2}$
instead of $f_{\rho 2}$ and $f_{\omega 2}$, yielding
\begin{eqnarray}
  {\left( \kappa_{\rho 2} \right)}_{\rm QM} = -1.5 \, &\mbox{vs}& 
  {\left( \kappa_{\rho 2} \right)}_{\rm HADS} = -1.1 \pm 0.2  \, , 
\end{eqnarray}
plus identical predictions for $\kappa_{\omega 2}$.
In this case the difference is bigger,
but both set of values remain compatible.
It is important to notice there that the HADS predictions are expected
to be subjected to a sizeable error of
$\Lambda_{QCD} / (m_Q v) \sim 30-40 \%$ in the charm sector
(instead of the standard $\Lambda_{\rm QCD} / m_Q \sim 10-15 \%$
for HQSS)~\cite{Savage:1990di,Guo:2013xga},
where $v$ is the velocity of the heavy quark
in the $\Xi_{cc}$ baryon.
For the quark model predictions the situation is a bit more murky,
owing to its status as a model (as they usually lack
reliable error estimations).
We warn however that the apparent similarity of both set of predictions
is not necessarily due to a compatibility between the two models:
the choice that we have made for the couplings of the cascade
have also played a role.

\section{Predictions of Molecular States}
\label{sec:predictions}

Now we solve the Schr\"odinger equation
for the $H \Xi$ and ${\bar H} \Xi$ potentials
with the coupling constant choices we have made in the previous section.
For the cut-off in the form factor we will fix $\Lambda$
as to reproduce the $X(3872)$ in the isospin symmetric limit.
In this limit the $X(3872)$ is a $1^{++}$ $D\bar{D}^*$ molecule
with a binding energy of about $4\,{\rm MeV}$,
which corresponds to a binding energy of about $0\,{\rm MeV}$
if we consider isospin symmetry breaking in the masses of the charmed mesons.
With the choices of the couplings previously made,
we obtain the value $\Lambda = \Lambda_X \simeq 1.04 \,{\rm GeV}$.
For this cut-off the charmed meson - cascade molecules do not bind
but are prettry close to binding.
The $J={\frac{1}{2}}^{-}$ $D^* \Xi$ system is the most attractive case.
It binds for $\Lambda \geq 1.05\,{\rm GeV}$,
which is just a tiny fraction above $\Lambda_X$.
Concrete calculations indicate a scattering length of $a_0 = -18.7\,{\rm fm}$,
which is indeed larger than any other scale in the system.
For the other two configurations of the charmed meson - cascade system
we find that the $J={\frac{1}{2}}^{-}$ $D \Xi$ and
${\frac{3}{2}}^{-}$ $D^* \Xi$ scattering length
is $a_0 = -1.8\,{\rm fm}$ in both cases,
indicating a moderate degree of attraction.

\begin{figure}[!th]
\centering
\begin{overpic}[scale=.3]{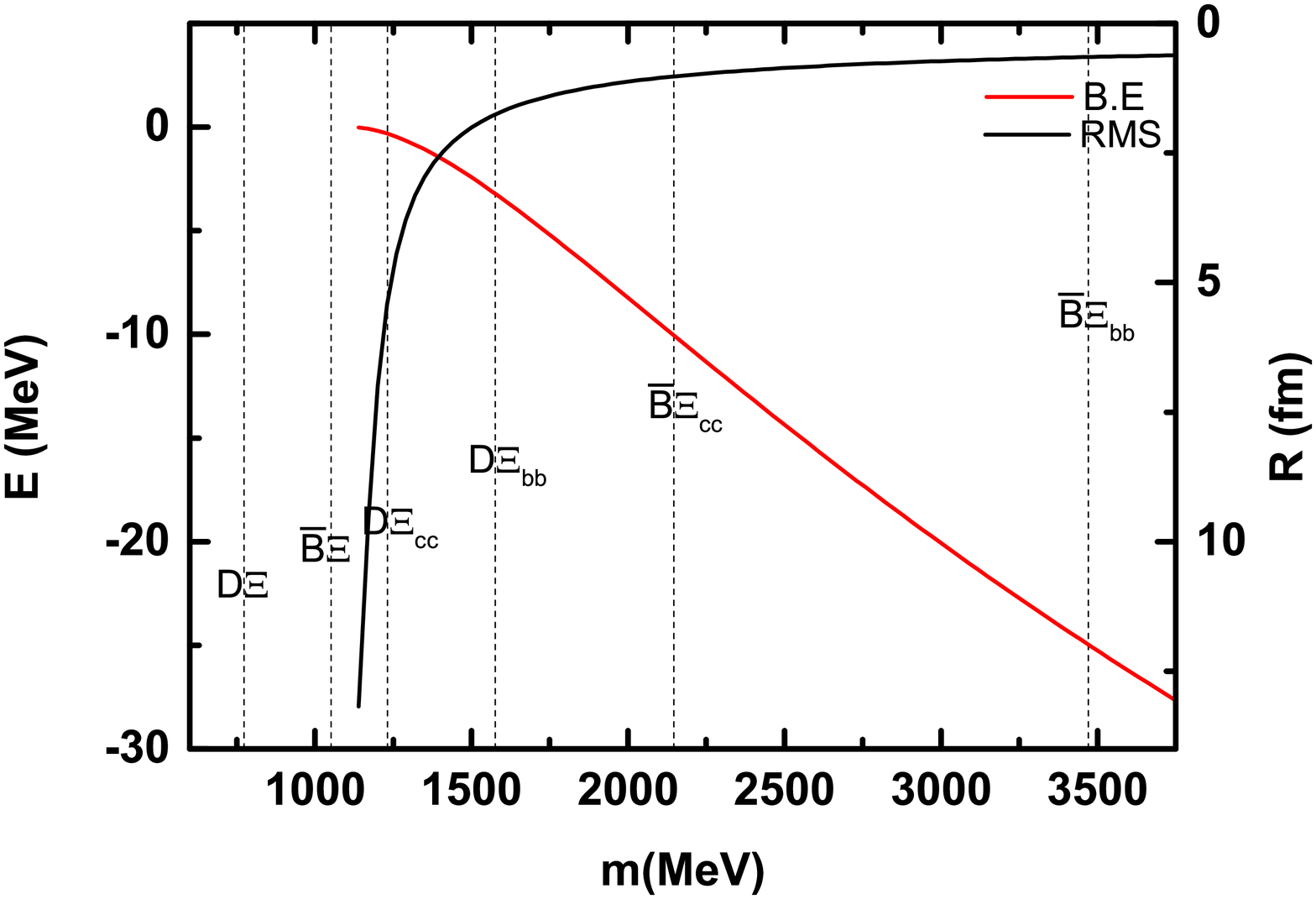}
\end{overpic}
\begin{overpic}[scale=.3]{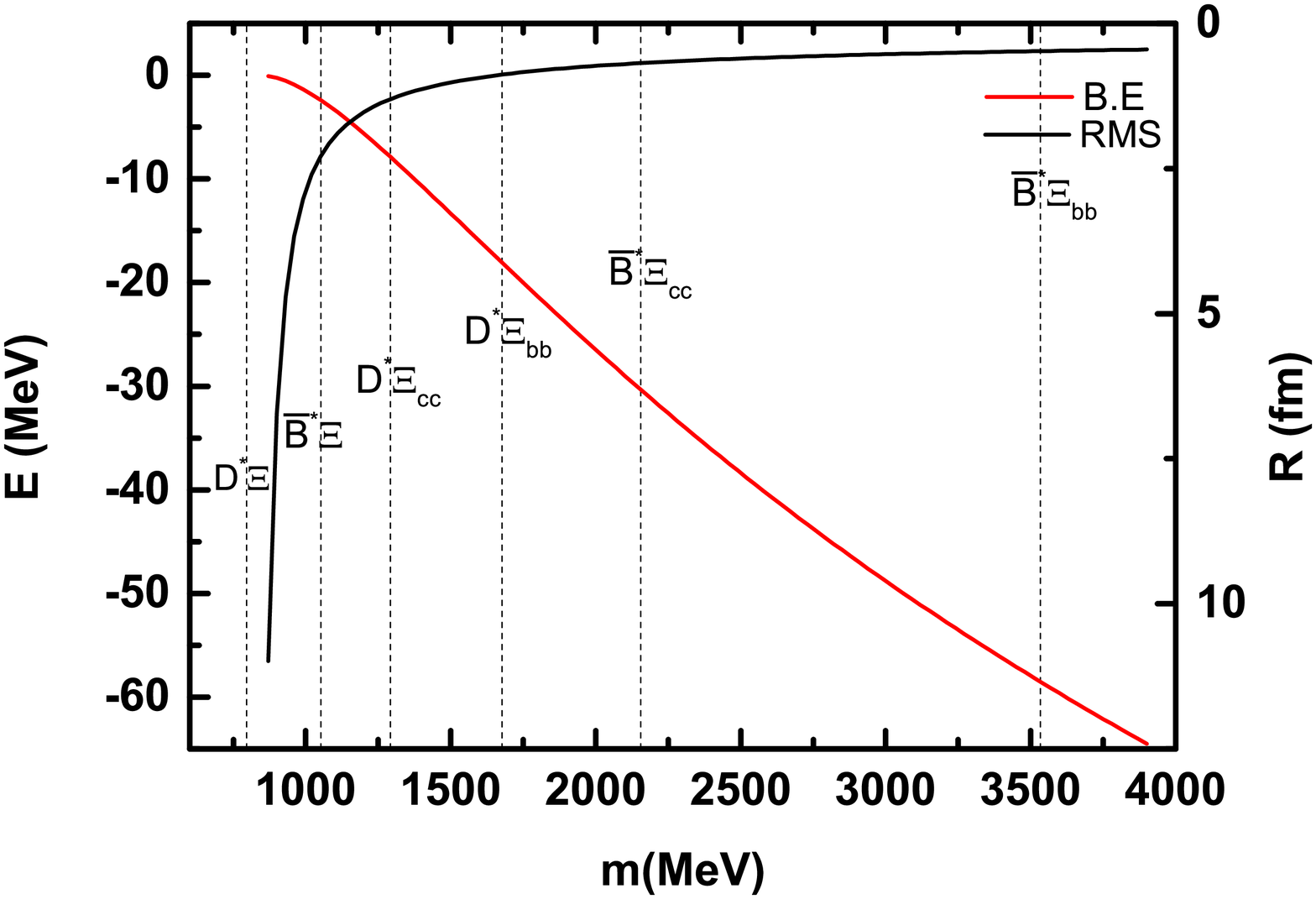}
\end{overpic}
\begin{overpic}[scale=.3]{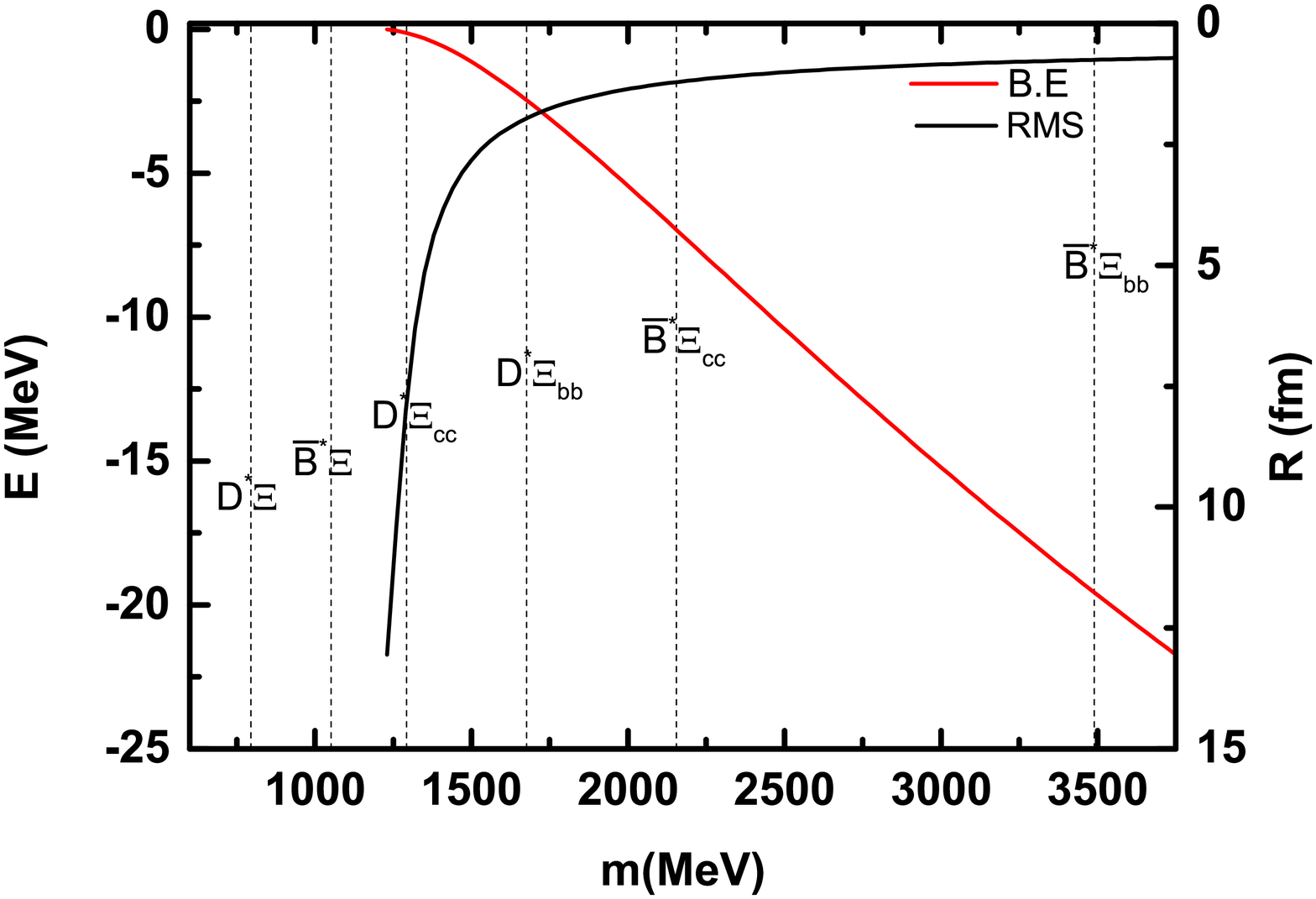}
\end{overpic}
\caption{
  Binding energies (red line) and root mean square radii (black line)
  for the $D\Xi$ and $D^*\Xi$ molecules depending on the reduced mass.
  The upper, middle and lower panel correspond to the $\frac{1}{2}^{-}$
  $D\Xi$/$D\Xi_{QQ}$, $\frac{1}{2}^{-}$ $D^*\Xi$/$D^*\Xi_{QQ}$ and
  $\frac{3}{2}^{-}$ $D^*\Xi$/$D^*\Xi_{QQ}$ molecules respectively.
  The vertical dotted lines indicate the reduced masses of the different
  molecules considered.
  The calculations are made with a form factor cut-off
  $\Lambda = \Lambda_X = 1.04\,{\rm GeV}$, which corresponds to the cut-off
  for which the $X(3872)$ is reproduced in the OBE model we use.
}~\label{Xi1}
\end{figure}

This attraction will become able to bind if we increase
the reduced mass of the system.
For the $\frac{1}{2}^{-}$ $\bar{B}^* \Xi$ molecule,
the bottom counterpart of the ${\frac{1}{2}}^{-}$ $D^* \Xi$,
binding happens at $B = 2.9\,{\rm MeV}$.
Meanwhile for the $\frac{1}{2}^{-}$ $\bar{B} \Xi$ and
$\frac{1}{2}^{-}$ $\bar{B}^* \Xi$ systems the scattering lengths
are expected to be large, $a_0 = -15.1\,{\rm fm}$
and $-7.2\,{\rm fm}$ respectively.
Owing to our choice of parameters for the standard cascade, we can basically
extend the present calculation to doubly heavy baryons by simply changing
the reduced mass in the calculations.
Indeed if we consider the charmed meson - doubly charmed baryon molecules,
we find the binding energy of ${\frac{1}{2}}^{-}$ $D^* \Xi_{cc}$ molecule
to be $B = 8.7\,{\rm MeV}$, while for the ${\frac{1}{2}}^{-}$ $D \Xi_{cc}$
and ${\frac{3}{2}}^{-}$ $D^* \Xi_{cc}$ molecules $B = 0.3\,{\rm MeV}$
and $0.2\,{\rm MeV}$ respectively.
As happened with the ${\frac{1}{2}}^{-}$ $D^* \Xi$ system,
the ${\frac{1}{2}}^{-}$ $D \Xi_{cc}$ and ${\frac{3}{2}}^{-}$ $D^* \Xi_{cc}$
molecules are close to the unitary limit,
where their scattering lengths are $a_0 = 7.6\,{\rm fm}$ and $10.1\,{\rm fm}$
respectively.
In Fig.~\ref{Xi1} we plot the dependence of the binding energy
on the reduced mass for the different $J^P$ molecular configurations,
where we also indicate the location of the thresholds.
For comparison purposes we also include the mean square radius of
these molecules in Fig.~\ref{Xi1}, which is important to determine
whether they are actual bound states (i.e. with a size bigger
than its components) or more compact objects requiring
a different type of treatment.

The previous numbers are subject to theoretical errors, which are however
not easy to estimate because the OBE potential is a model after all.
The coupling constants in the potential are in principle amenable
to error estimations.
The values we take for $g_{\omega 1}$ and $g_{\omega 2}$ are derived
from SU(3) and the OZI rule and hence expected to have
an uncertainty of about $20\,\%$.
For the couplings derived from the quark model, $g_{\sigma 1}$ and $g_{\sigma 2}$,
it is less clear which uncertainty to adscribe them,
but probably a $30\,\%$ could be a good guess.
For $g_{\rho 1}$ and $g_{\rho 2}$, the error depends on how much do we expect
the KSFR  relation to fail,
but probably a $10\,\%$ is enough.
The axial coupling and its error $g_1$ are known experimentally,
while for $g_2$ the error is again determined from the quark model.
However the independent variation of each of the couplings followed
by the subsequent addition of the errors in quadrature is cumbersome
because of the large number of parameters to vary (not to mention
that it is not so easy to determine the error of all of them).
We find instead much more convenient to simply assume a global uncertainty
for the $D$/$D^*$ and $\Xi$ couplings in the following way
\begin{eqnarray}
  g_{M1} (1 \pm \Delta_1) \quad \mbox{and} \quad g_{M2} (1 \pm \Delta_2) \, ,
\end{eqnarray}
where $M$ stands for the $\pi$, $\sigma$, $\rho$ and $\omega$ mesons and
with $\Delta_1$ and $\Delta_2$ the relative error we expect
in each of the vertices.
If we assume all the couplings to vary in the same direction,
i.e. correlated errors, then the outcome is that there is
an overall uncertainty in the $X(3872)$ and $D\Xi$/$D^*\Xi$ potentials.
In this picture for a potential with a vertex of type $i$ and $j$
we will assign the error
\begin{eqnarray}
  V_{ij} \, (1 \pm \Delta_j) \, (1 \pm \Delta_j) \, , 
\end{eqnarray}
where vertex type $1$ refers to a $D$/$D^*$ and vertex type $2$ to a cascade.
We will assume the uncertainties on vertex type $1$ and $2$ to be uncorrelated.
In the previous notation, the $X(3872)$ potential will be $V_{11}$ and
the $D\Xi$/$D^*\Xi$ potential will be $V_{12}$.
Besides it is important to stress that the role of the couplings
to the charmed mesons and the cascade play a fundamentally
different role in the calculations.
We are using the $X(3872)$ and hence the couplings of the charmed mesons as a
way to fix the unkown parameter $\Lambda$ in the OBE model,
i.e. as a sort of {\it renormalization condition}.
That is, a change in the charmed meson vertex piece of the potential
entails a change in $\Lambda_X$ from which to redo the predictions
of the binding energy:
\begin{eqnarray}
  V_{11}' = V_{11}\,(1 \pm \Delta_1)^2 \to \Lambda_X' \to B_{D\Xi}'
\end{eqnarray}
where $B_{D\Xi}'$ is the binding energy for $\Lambda_X'$,
where the parameters for vertex $1$ in $V_{12}$
have to change congruently as how they change in $V_{11}$.
After this the error of the cascade baryon vertex $(1 \pm \Delta_2)$
should be added in quadrature.
If we follow this procedure and assume a global $\Delta_1 = 0.15$,
i.e. a $30\%$ global error in the $X(3872)$ potential,
we get $\Lambda_X = 1.04^{+0.18}_{-0.10}\,{\rm MeV}$.
If we apply this idea to vertex $2$ with $\Delta_2 = 0.15$,
the predictions and uncertainties for inverse of the $D\Xi$,
$D^*\Xi$, $\bar{B}\Xi$ and $\bar{B}^*\Xi$ scattering lengths
can be found in Table~\ref{tab:a0}.
Notice the choice of the inverse scattering length: the reason is that
the scattering length diverges and then changes sign
when there is a bound state.
Its inverse however changes smoothly, and hence the choice.
Actually the uncertainties are still compatible
with the existence of bound states in the $D\Xi$ and $D^*\Xi$ systems.
For the $P \Xi_{QQ}$ and $P^* \Xi_{QQ}$ bound states, their binding energies
and uncertaintie can be consulted in Table~\ref{tab:binding}.

\begin{table}[!t]
\begin{tabular}{|ccc|ccc|}
\hline\hline
state  & $I\,(J^{P})$  & $\frac{1}{a_0}$ (MeV) &
state  & $I\,(J^{P})$  & $\frac{1}{a_0}$ (MeV) \\
  \hline
  $D \Xi$ & $0$($\frac{1}{2}^{-}$) & $-110^{+110}_{-120}$ &
  $\bar{B} \Xi$ & $0$($\frac{1}{2}^{-}$) & $-10^{+80}_{-90}$
  \\
  $D^* \Xi$ & $0$($\frac{1}{2}^{-}$) & $-10^{+70}_{-70}$ &
  $\bar{B}^* \Xi$ & $0$($\frac{1}{2}^{-}$) & $+70^{+60}_{-60}$ 
  \\
  $D^* \Xi$ & $0$($\frac{3}{2}^{-}$) & $-110^{+110}_{-120}$ &
  $\bar{B}^* \Xi$ & $0$($\frac{3}{2}^{-}$) & $-20^{+100}_{-100}$
  \\
  \hline\hline 
\end{tabular}
\caption{
  Inverse of the scattering length for the $D\Xi$, $D^*\Xi$ systems
  and their bottom counterparts $\bar{B}\Xi$ and $\bar{B}^*\Xi$ and.
  They are deduced from the condition of reproducing the $X(3872)$ pole
  in the $D\bar{D}^*$ potential and by assuming a $15\%$ error
  in the $D$/$D^*$ and $\Xi$ vertices (where the same applies
  to the $\bar{B}\Xi$ and $\bar{B}^*\Xi$ systems).
  }
\label{tab:a0}
\end{table}

\begin{table}[!h]
\begin{tabular}{|ccc|ccc|}
\hline\hline
state  & $I\,(J^{P})$  & B (MeV) &
state  & $I\,(J^{P})$  & B (MeV) \\
  \hline
  $D \Xi_{cc}$ & $0$($\frac{1}{2}^{-}$) & $0^{+8}_{\dagger}$ &
    $\bar{B} \Xi_{cc}$ & $0$($\frac{1}{2}^{-}$) & $10^{+25}_{-9}$ \\
  $D^* \Xi_{cc}$ & $0$($\frac{1}{2}^{-}$) & $9^{+15}_{-8}$ &
    $\bar{B}^* \Xi_{cc}$ & $0$($\frac{1}{2}^{-}$) & $32^{+27}_{-17}$ \\
  $D^* \Xi_{cc}$ & $0$($\frac{3}{2}^{-}$) & $0^{+9}_{\dagger}$ &
    $\bar{B}^* \Xi_{cc}$ & $0$($\frac{3}{2}^{-}$) & $8^{+28}_{\dagger}$ \\
  \hline
  $D \Xi_{bb}$ & $0$($\frac{1}{2}^{-}$) & $3^{+15}_{\dagger}$ &
    $\bar{B} \Xi_{cc}$ & $0$($\frac{1}{2}^{-}$) & $25^{+39}_{-18}$ \\
  $D^* \Xi_{bb}$ & $0$($\frac{1}{2}^{-}$) & $19^{+21}_{-13}$ &
    $\bar{B}^* \Xi_{cc}$ & $0$($\frac{1}{2}^{-}$) & $60^{+36}_{-25}$ \\
  $D^* \Xi_{bb}$ & $0$($\frac{3}{2}^{-}$) & $3^{+16}_{\dagger}$ &
    $\bar{B}^* \Xi_{cc}$ & $0$($\frac{3}{2}^{-}$) & $21^{+47}_{-16}$ \\
  \hline\hline 
\end{tabular}
\caption{
  Binding energies for the molecular states considered in this work
  from the condition that the form factor cut-off is chosen to
  reproduce the $X(3872)$ pole.
  The ${\dagger}$ symbol is used to indicate that
  the error is large enough as to unbind the system.
  }
\label{tab:binding}
\end{table}

We stress that the previous conclusions are derived
from the hypothesis that the $X(3872)$ is molecular
at the distances in which the OBE model applies.
Besides the circumstantial fact that the $X(3872)$ is located close
to the $D^0 \bar{D}^{0*}$ threshold, the most convincing evidence
that the $X$ is molecular is the ratio of its isospin breaking decays
$\Gamma(X \to J/\Psi 2 \pi)$ and
$\Gamma(X \to J/\Psi 3 \pi)$~\cite{Choi:2011fc}.
It is relatively easy to explain this branching ratio within
the molecular picture~\cite{Gamermann:2009fv,Gamermann:2009uq},
but not so if the $X(3872)$ is a compact charmonium-like
state~\cite{Hanhart:2011tn}.
However the radiative decays $\Gamma(X \to J/\Psi \gamma)$ and
$\Gamma(X \to \Psi(2S) \gamma)$~\cite{Aaij:2014ala} offer
a different perspective of the $X(3872)$, as they are
difficult to explain without the existence of $c\bar{c}$
components in the $X$ wave function~\cite{Swanson:2004pp}.
The $c\bar{c}$ component necessary to successfully explain
the radiative components is small~\cite{Dong:2009uf},
but could nonetheless represents a source of
systematic uncertainty for our predictions.
In this regard it has been shown that the $c\bar{c}$ short range components
do not necessarily affect the long range picture of the $X(3872)$ as a
hadronic molecule~\cite{Guo:2014taa}.
This means that while these components are important at short distances,
they are probably heavily suppressed at long distances.
This is consistent with the approximation we are using here
that the $X(3872)$ is molecular. Thus we do not expect
significant corrections to our predictions.

\section{Summary}
\label{sec:summary}

In this work we have considered the $D \Xi$ and $D^* \Xi$ systems
from the point of view of the OBE model and looked for possible
molecular states and their locations.
The OBE potential is traditionally regularized with a form factor and a cut-off.
The cut-off is expected to be in the $1\,{\rm GeV}$ range,
but there is a considerable uncertainty with respect to its value
that translates into wildly different predictions.
To reduce this uncertainty we have determined the cut-off from the condition
of reproducing the $X(3872)$ as a $1^{++}$ $D^* \bar{D}$ molecular state,
yielding $\Lambda = 1.04^{+0.18}_{-0.10}\,{\rm GeV}$.
With this we find that the $\frac{1}{2}^-$ $D^* \Xi$ state is on the verge
of binding, which translates into an unusually large scattering length
of $-18.7\,{\rm fm}$.
If we consider the theoretical uncertainties it turns out that this molecule
might very well bind, with the probability of binding being slightly smaller
than the probability of not binding.
This molecule has also been predicted in Ref.~\cite{Debastiani:2017ewu}.
The $\frac{1}{2}^-$ $D \Xi$ and $\frac{3}{2}^-$ $D^* \Xi$ are unlikely to bind,
though their interaction is indeed attractive as can be deduced from their
scattering lenghts.
As a consequence the interpretation of the $\Omega_c(3188)$ enhancement
as a $D \Xi$ bound state~\cite{Wang:2017smo} is disfavoured.
The conclusion about a possible $\frac{3}{2}^-$ $D^* \Xi$ molecule is
also different from Ref.~\cite{Debastiani:2017ewu},
where it is predicted, but the previous work includes a series of
coupled channels that increase the binding by a small amount.
Besides, there is the possibility that these two molecules bind
within the uncertainties of our model.

The previous findings can be easily extended to systems in which
the $D$ and $D^*$ are substituted by a $B$ and $B^*$ or where
instead of the cascade $\Xi$ we have a doubly heavy baryon
$\Xi_{cc}$ or $\Xi_{bb}$.
These systems have a large reduced mass and are thus more likely to bind.
For the $\Omega_b$-like molecular state, $\bar{B} \Xi$ and $\bar{B}^* \Xi$,
we find that the $\frac{1}{2}^{-}$ $\bar{B}^* \Xi$ binds withing the errors
of the present model, where this state have also been predicted
in \cite{Liang:2017ejq}.
The other two configurations might bind as well,
but this is contingent on the uncertainties.
For the triply heavy molecules
we find the $\frac{1}{2}^{-}$ $P^* \Xi_{QQ}$ system
to be the most attractive, binding in all cases.
The other two configurations --- $\frac{1}{2}^{-}$ $P \Xi_{QQ}$ and
$\frac{3}{2}^{-}$ $P^* \Xi_{QQ}$ --- are less attractive.
For the triply charmed pentaquark case the previous two configurations are
probably close to the unitary limit, where their central values
for the binding energy are $0.3$ and $0.2\,{\rm MeV}$
respectively.
For the triply bottom pentaquarks, all configurations bind
within the theoretical uncertainties of the present model.
Triply heavy pentaquarks have been considered previously in the literature.
In Ref.~\cite{Guo:2013xga}
HADS is applied to the $X(3872)$ as a $D^*\bar{D}$ molecule to deduce
the existence of possible $P \Xi_{QQ}$, $P^* \Xi_{QQ}$, $P \Xi_{QQ}^*$
and $P^* \Xi_{QQ}^*$ bound states.
From the $X(3872)$ the existence of isoscalar $\frac{5}{2}^{-}$ $P^* \Xi_{QQ}^*$
pentaquark-like molecules can indeed be deduced,
while for the other $J^P$ combinations the information that can be obtained
from the $X(3872)$ is insufficient to predict more states.
In this context the OBE model provides a phenomenological estimation of
this missing dynamics, which allows us to fully explore
the $P \Xi_{QQ}$ and $P^* \Xi_{QQ}$ cases.

\section*{Acknowledgments}
We thank Eulogio Oset for his careful reading of this manuscript.
This work is partly supported by the National Natural Science Foundation of China under Grants No.11522539, 11735003,
the fundamental Research Funds for the Central Universities,
the Youth Innovation Promotion Association CAS (No. 2016367)
and the Thousand Talents Plan for Young Professionals.

\appendix

\section{Couplings in the Quark Model}

Here we present how to compute the couplings of the light mesons
to different hadrons in the quark model.
At the quark level the Lagrangians describing quark interactions
with light mesons can be written as~\cite{Riska:2000gd}:
\begin{eqnarray}
  \mathcal{L}_{M qq}&=&
  g_{\pi qq}\,(\bar{u}i\gamma_{5}u\pi^{0}-\bar{d}i\gamma_{5}d\pi^{0})
  \\ \nonumber &+&
  g_{\omega qq}\,
  (\bar{u}\gamma_{\mu}u\omega^{\mu}+\bar{d}\gamma_{u}d\omega^{\mu}) \label{Lq}
  \\  \nonumber
  &+&g_{\rho qq}\,
  (\bar{u}\gamma_{\mu}u\rho^{0\mu}-\bar{d}\gamma_{\mu}d\rho^{0\mu})\\
  \nonumber &+&g_{\sigma qq}\,(\bar{u}u\sigma+\bar{d}d\sigma),
\end{eqnarray}
where $g_{\pi qq}$, $g_{\rho qq}$, $g_{\omega qq}$ and $g_{\sigma qq}$ are
the couplings of the light $q = u, d$ quarks to the light mesons.
The lagrangian above assumes a non-derivative coupling with the pion,
which is non chirally symmetric.
This is inconsequential as the chiral derivative coupling leads to
the same tree level description, i.e. the same one pion exchange potential.
To obtain the relations between $g_{M qq}$ and $g_{M hh}$, where the later
refers to the couplings with an arbitrary hadron $h$, we require
the interaction vertices calculated at the quark and
hadron levels be the same, i.e.,
\begin{equation}
\langle h,\vec{s}\,| \,\mathcal{L}_{M hh}\, |h,\vec{s}\rangle\equiv
\langle h,\vec{s}\,| \,\mathcal{L}_{M qq}\, |h,\vec{s}\rangle,
\end{equation}
where $H$ denotes a hadron, $\vec{s}$ its spin and $\mathcal{L}_{M HH}$ is
the OBE lagrangian for the hadron $H$.
For instance, let us consider the case of the coupling of nucleon
and the cascade to the pion
\begin{eqnarray}
  \langle p\uparrow |\,\mathcal{L}_{\pi NN}|p\uparrow\rangle
  &=&
  \frac{g_{\pi  N N}}{m_N}\,q_3\, , \\
  \langle \Xi^{0} \uparrow | \mathcal{L}_{\pi \Xi \Xi}|
  \Xi^{0}\uparrow\rangle&=& \frac{g_{\pi \Xi\Xi}}{m_{\Xi}}\,q_3 \, ,
\end{eqnarray}
where $q$ refers to the momentum of the pion.
We can directly compare the previous matrix elements to the ones we obtain
from the SU(6) quark model wave functions~\cite{Riska:2000gd} yielding
\begin{eqnarray}
  \langle p\uparrow | \mathcal{L}_{\pi qq} |p\uparrow \rangle
  &=&\frac{5}{3} \frac{g_{\pi q q}}{m_q}\,q_3\, , \\
  \langle \Xi^{0}\uparrow | \mathcal{L}_{\pi qq} | \Xi^{0}\uparrow\rangle
  &=&-\frac{1}{3}\,\frac{g_{\pi q q}}{m_q}\,q_3 \, .
\end{eqnarray}
A direct comparison gives us the relation between $g_{\pi \Xi\Xi}$ and $g_{\pi NN}$:
\begin{eqnarray}
g_{\pi \Xi\Xi}=-\frac{1}{5}\frac{m_{\Xi}}{m_{N}}g_{\pi NN} \, .
\end{eqnarray}
Repeating this procedure for the other light mesons, we obtain
\begin{eqnarray}
 g_{\sigma \Xi \Xi} &=& \frac{1}{3}\,g_{\sigma NN} \, ,  \\
 g_{\omega \Xi \Xi} &=& \frac{1}{3}g_{\omega NN} \, , \\
 (g_{\omega\Xi\Xi}+f_{\omega \Xi\Xi}) &=&
 -\frac{1}{3}\frac{m_{\Xi}}{m_{N}}(g_{\omega  N N}+f_{\omega N N}) \, , \\
 g_{\rho \Xi \Xi} &=& g_{\rho N N} \, , \\
 (g_{\rho \Xi \Xi}+f_{\rho \Xi \Xi}) &=& -\frac{1}{5}
 \frac{m_{\Xi}}{m_{N}}(g_{\rho NN}+f_{\rho NN}) \, . 
\end{eqnarray}
Thus we can relate the nucleon coupling constants with the ones for the cascade
or with the ones for other hadrons.

%\bibliography{DXi-3320.bib}

\begin{thebibliography}{79}
\expandafter\ifx\csname natexlab\endcsname\relax\def\natexlab#1{#1}\fi
\expandafter\ifx\csname bibnamefont\endcsname\relax
  \def\bibnamefont#1{#1}\fi
\expandafter\ifx\csname bibfnamefont\endcsname\relax
  \def\bibfnamefont#1{#1}\fi
\expandafter\ifx\csname citenamefont\endcsname\relax
  \def\citenamefont#1{#1}\fi
\expandafter\ifx\csname url\endcsname\relax
  \def\url#1{\texttt{#1}}\fi
\expandafter\ifx\csname urlprefix\endcsname\relax\def\urlprefix{URL }\fi
\providecommand{\bibinfo}[2]{#2}
\providecommand{\eprint}[2][]{\url{#2}}

\bibitem[{\citenamefont{Choi et~al.}(2003)}]{Choi:2003ue}
\bibinfo{author}{\bibfnamefont{S.~K.} \bibnamefont{Choi}} \bibnamefont{et~al.}
  (\bibinfo{collaboration}{Belle}), \bibinfo{journal}{Phys. Rev. Lett.}
  \textbf{\bibinfo{volume}{91}}, \bibinfo{pages}{262001}
  (\bibinfo{year}{2003}), \eprint{hep-ex/0309032}.

\bibitem[{\citenamefont{Guo et~al.}(2018)\citenamefont{Guo, Hanhart, Meißner,
  Wang, Zhao, and Zou}}]{Guo:2017jvc}
\bibinfo{author}{\bibfnamefont{F.-K.} \bibnamefont{Guo}},
  \bibinfo{author}{\bibfnamefont{C.}~\bibnamefont{Hanhart}},
  \bibinfo{author}{\bibfnamefont{U.-G.} \bibnamefont{Meißner}},
  \bibinfo{author}{\bibfnamefont{Q.}~\bibnamefont{Wang}},
  \bibinfo{author}{\bibfnamefont{Q.}~\bibnamefont{Zhao}}, \bibnamefont{and}
  \bibinfo{author}{\bibfnamefont{B.-S.} \bibnamefont{Zou}},
  \bibinfo{journal}{Rev. Mod. Phys.} \textbf{\bibinfo{volume}{90}},
  \bibinfo{pages}{015004} (\bibinfo{year}{2018}), \eprint{1705.00141}.

\bibitem[{\citenamefont{Ablikim et~al.}(2013{\natexlab{a}})}]{Ablikim:2013mio}
\bibinfo{author}{\bibfnamefont{M.}~\bibnamefont{Ablikim}} \bibnamefont{et~al.}
  (\bibinfo{collaboration}{BESIII}), \bibinfo{journal}{Phys. Rev. Lett.}
  \textbf{\bibinfo{volume}{110}}, \bibinfo{pages}{252001}
  (\bibinfo{year}{2013}{\natexlab{a}}), \eprint{1303.5949}.

\bibitem[{\citenamefont{Liu et~al.}(2013)}]{Liu:2013dau}
\bibinfo{author}{\bibfnamefont{Z.~Q.} \bibnamefont{Liu}} \bibnamefont{et~al.}
  (\bibinfo{collaboration}{Belle}), \bibinfo{journal}{Phys. Rev. Lett.}
  \textbf{\bibinfo{volume}{110}}, \bibinfo{pages}{252002}
  (\bibinfo{year}{2013}), \eprint{1304.0121}.

\bibitem[{\citenamefont{Ablikim et~al.}(2013{\natexlab{b}})}]{Ablikim:2013wzq}
\bibinfo{author}{\bibfnamefont{M.}~\bibnamefont{Ablikim}} \bibnamefont{et~al.}
  (\bibinfo{collaboration}{BESIII}), \bibinfo{journal}{Phys. Rev. Lett.}
  \textbf{\bibinfo{volume}{111}}, \bibinfo{pages}{242001}
  (\bibinfo{year}{2013}{\natexlab{b}}), \eprint{1309.1896}.

\bibitem[{\citenamefont{Ablikim et~al.}(2014)}]{Ablikim:2014dxl}
\bibinfo{author}{\bibfnamefont{M.}~\bibnamefont{Ablikim}} \bibnamefont{et~al.}
  (\bibinfo{collaboration}{BESIII}), \bibinfo{journal}{Phys. Rev. Lett.}
  \textbf{\bibinfo{volume}{113}}, \bibinfo{pages}{212002}
  (\bibinfo{year}{2014}), \eprint{1409.6577}.

\bibitem[{\citenamefont{Bondar et~al.}(2012)}]{Belle:2011aa}
\bibinfo{author}{\bibfnamefont{A.}~\bibnamefont{Bondar}} \bibnamefont{et~al.}
  (\bibinfo{collaboration}{Belle}), \bibinfo{journal}{Phys. Rev. Lett.}
  \textbf{\bibinfo{volume}{108}}, \bibinfo{pages}{122001}
  (\bibinfo{year}{2012}), \eprint{1110.2251}.

\bibitem[{\citenamefont{Garmash et~al.}(2015)}]{Garmash:2014dhx}
\bibinfo{author}{\bibfnamefont{A.}~\bibnamefont{Garmash}} \bibnamefont{et~al.}
  (\bibinfo{collaboration}{Belle}), \bibinfo{journal}{Phys. Rev.}
  \textbf{\bibinfo{volume}{D91}}, \bibinfo{pages}{072003}
  (\bibinfo{year}{2015}), \eprint{1403.0992}.

\bibitem[{\citenamefont{Aaij et~al.}(2015)}]{Aaij:2015tga}
\bibinfo{author}{\bibfnamefont{R.}~\bibnamefont{Aaij}} \bibnamefont{et~al.}
  (\bibinfo{collaboration}{LHCb}), \bibinfo{journal}{Phys. Rev. Lett.}
  \textbf{\bibinfo{volume}{115}}, \bibinfo{pages}{072001}
  (\bibinfo{year}{2015}), \eprint{1507.03414}.

\bibitem[{\citenamefont{Voloshin and Okun}(1976)}]{Voloshin:1976ap}
\bibinfo{author}{\bibfnamefont{M.}~\bibnamefont{Voloshin}} \bibnamefont{and}
  \bibinfo{author}{\bibfnamefont{L.}~\bibnamefont{Okun}},
  \bibinfo{journal}{JETP Lett.} \textbf{\bibinfo{volume}{23}},
  \bibinfo{pages}{333} (\bibinfo{year}{1976}).

\bibitem[{\citenamefont{De~Rujula et~al.}(1977)\citenamefont{De~Rujula, Georgi,
  and Glashow}}]{DeRujula:1976zlg}
\bibinfo{author}{\bibfnamefont{A.}~\bibnamefont{De~Rujula}},
  \bibinfo{author}{\bibfnamefont{H.}~\bibnamefont{Georgi}}, \bibnamefont{and}
  \bibinfo{author}{\bibfnamefont{S.~L.} \bibnamefont{Glashow}},
  \bibinfo{journal}{Phys. Rev. Lett.} \textbf{\bibinfo{volume}{38}},
  \bibinfo{pages}{317} (\bibinfo{year}{1977}).

\bibitem[{\citenamefont{Tornqvist}(1991)}]{Tornqvist:1991ks}
\bibinfo{author}{\bibfnamefont{N.~A.} \bibnamefont{Tornqvist}},
  \bibinfo{journal}{Phys.Rev.Lett.} \textbf{\bibinfo{volume}{67}},
  \bibinfo{pages}{556} (\bibinfo{year}{1991}), \bibinfo{note}{revised version}.

\bibitem[{\citenamefont{Tornqvist}(1994)}]{Tornqvist:1993ng}
\bibinfo{author}{\bibfnamefont{N.~A.} \bibnamefont{Tornqvist}},
  \bibinfo{journal}{Z.Phys.} \textbf{\bibinfo{volume}{C61}},
  \bibinfo{pages}{525} (\bibinfo{year}{1994}), \eprint{hep-ph/9310247}.

\bibitem[{\citenamefont{Manohar and Wise}(1993)}]{Manohar:1992nd}
\bibinfo{author}{\bibfnamefont{A.~V.} \bibnamefont{Manohar}} \bibnamefont{and}
  \bibinfo{author}{\bibfnamefont{M.~B.} \bibnamefont{Wise}},
  \bibinfo{journal}{Nucl.Phys.} \textbf{\bibinfo{volume}{B399}},
  \bibinfo{pages}{17} (\bibinfo{year}{1993}), \eprint{hep-ph/9212236}.

\bibitem[{\citenamefont{Ericson and Karl}(1993)}]{Ericson:1993wy}
\bibinfo{author}{\bibfnamefont{T.~E.~O.} \bibnamefont{Ericson}}
  \bibnamefont{and} \bibinfo{author}{\bibfnamefont{G.}~\bibnamefont{Karl}},
  \bibinfo{journal}{Phys.Lett.} \textbf{\bibinfo{volume}{B309}},
  \bibinfo{pages}{426} (\bibinfo{year}{1993}).

\bibitem[{\citenamefont{Weinberg}(1965)}]{Weinberg:1965zz}
\bibinfo{author}{\bibfnamefont{S.}~\bibnamefont{Weinberg}},
  \bibinfo{journal}{Phys. Rev.} \textbf{\bibinfo{volume}{137}},
  \bibinfo{pages}{B672} (\bibinfo{year}{1965}).

\bibitem[{\citenamefont{Liu et~al.}(2008)\citenamefont{Liu, Liu, Deng, and
  Zhu}}]{Liu:2008fh}
\bibinfo{author}{\bibfnamefont{Y.-R.} \bibnamefont{Liu}},
  \bibinfo{author}{\bibfnamefont{X.}~\bibnamefont{Liu}},
  \bibinfo{author}{\bibfnamefont{W.-Z.} \bibnamefont{Deng}}, \bibnamefont{and}
  \bibinfo{author}{\bibfnamefont{S.-L.} \bibnamefont{Zhu}},
  \bibinfo{journal}{Eur. Phys. J.} \textbf{\bibinfo{volume}{C56}},
  \bibinfo{pages}{63} (\bibinfo{year}{2008}), \eprint{0801.3540}.

\bibitem[{\citenamefont{Liu et~al.}(2009)\citenamefont{Liu, Luo, Liu, and
  Zhu}}]{Liu:2008tn}
\bibinfo{author}{\bibfnamefont{X.}~\bibnamefont{Liu}},
  \bibinfo{author}{\bibfnamefont{Z.-G.} \bibnamefont{Luo}},
  \bibinfo{author}{\bibfnamefont{Y.-R.} \bibnamefont{Liu}}, \bibnamefont{and}
  \bibinfo{author}{\bibfnamefont{S.-L.} \bibnamefont{Zhu}},
  \bibinfo{journal}{Eur. Phys. J.} \textbf{\bibinfo{volume}{C61}},
  \bibinfo{pages}{411} (\bibinfo{year}{2009}), \eprint{0808.0073}.

\bibitem[{\citenamefont{Sun et~al.}(2011)\citenamefont{Sun, He, Liu, Luo, and
  Zhu}}]{Sun:2011uh}
\bibinfo{author}{\bibfnamefont{Z.-F.} \bibnamefont{Sun}},
  \bibinfo{author}{\bibfnamefont{J.}~\bibnamefont{He}},
  \bibinfo{author}{\bibfnamefont{X.}~\bibnamefont{Liu}},
  \bibinfo{author}{\bibfnamefont{Z.-G.} \bibnamefont{Luo}}, \bibnamefont{and}
  \bibinfo{author}{\bibfnamefont{S.-L.} \bibnamefont{Zhu}},
  \bibinfo{journal}{Phys. Rev.} \textbf{\bibinfo{volume}{D84}},
  \bibinfo{pages}{054002} (\bibinfo{year}{2011}), \eprint{1106.2968}.

\bibitem[{\citenamefont{He}(2015)}]{He:2015mja}
\bibinfo{author}{\bibfnamefont{J.}~\bibnamefont{He}}, \bibinfo{journal}{Phys.
  Rev.} \textbf{\bibinfo{volume}{D92}}, \bibinfo{pages}{034004}
  (\bibinfo{year}{2015}), \eprint{1505.05379}.

\bibitem[{\citenamefont{Sun et~al.}(2012)\citenamefont{Sun, Luo, He, Liu, and
  Zhu}}]{Sun:2012zzd}
\bibinfo{author}{\bibfnamefont{Z.-F.} \bibnamefont{Sun}},
  \bibinfo{author}{\bibfnamefont{Z.-G.} \bibnamefont{Luo}},
  \bibinfo{author}{\bibfnamefont{J.}~\bibnamefont{He}},
  \bibinfo{author}{\bibfnamefont{X.}~\bibnamefont{Liu}}, \bibnamefont{and}
  \bibinfo{author}{\bibfnamefont{S.-L.} \bibnamefont{Zhu}},
  \bibinfo{journal}{Chin. Phys.} \textbf{\bibinfo{volume}{C36}},
  \bibinfo{pages}{194} (\bibinfo{year}{2012}).

\bibitem[{\citenamefont{Dias et~al.}(2015)\citenamefont{Dias, Aceti, and
  Oset}}]{Dias:2014pva}
\bibinfo{author}{\bibfnamefont{J.~M.} \bibnamefont{Dias}},
  \bibinfo{author}{\bibfnamefont{F.}~\bibnamefont{Aceti}}, \bibnamefont{and}
  \bibinfo{author}{\bibfnamefont{E.}~\bibnamefont{Oset}},
  \bibinfo{journal}{Phys. Rev.} \textbf{\bibinfo{volume}{D91}},
  \bibinfo{pages}{076001} (\bibinfo{year}{2015}), \eprint{1410.1785}.

\bibitem[{\citenamefont{Wang et~al.}(2018)\citenamefont{Wang, Baru, Filin,
  Hanhart, Nefediev, and Wynen}}]{Wang:2018jlv}
\bibinfo{author}{\bibfnamefont{Q.}~\bibnamefont{Wang}},
  \bibinfo{author}{\bibfnamefont{V.}~\bibnamefont{Baru}},
  \bibinfo{author}{\bibfnamefont{A.~A.} \bibnamefont{Filin}},
  \bibinfo{author}{\bibfnamefont{C.}~\bibnamefont{Hanhart}},
  \bibinfo{author}{\bibfnamefont{A.~V.} \bibnamefont{Nefediev}},
  \bibnamefont{and} \bibinfo{author}{\bibfnamefont{J.~L.} \bibnamefont{Wynen}}
  (\bibinfo{year}{2018}), \eprint{1805.07453}.

\bibitem[{\citenamefont{Chen et~al.}(2015{\natexlab{a}})\citenamefont{Chen,
  Liu, Li, and Zhu}}]{Chen:2015loa}
\bibinfo{author}{\bibfnamefont{R.}~\bibnamefont{Chen}},
  \bibinfo{author}{\bibfnamefont{X.}~\bibnamefont{Liu}},
  \bibinfo{author}{\bibfnamefont{X.-Q.} \bibnamefont{Li}}, \bibnamefont{and}
  \bibinfo{author}{\bibfnamefont{S.-L.} \bibnamefont{Zhu}},
  \bibinfo{journal}{Phys. Rev. Lett.} \textbf{\bibinfo{volume}{115}},
  \bibinfo{pages}{132002} (\bibinfo{year}{2015}{\natexlab{a}}),
  \eprint{1507.03704}.

\bibitem[{\citenamefont{Chen et~al.}(2015{\natexlab{b}})\citenamefont{Chen,
  Chen, Liu, Steele, and Zhu}}]{Chen:2015moa}
\bibinfo{author}{\bibfnamefont{H.-X.} \bibnamefont{Chen}},
  \bibinfo{author}{\bibfnamefont{W.}~\bibnamefont{Chen}},
  \bibinfo{author}{\bibfnamefont{X.}~\bibnamefont{Liu}},
  \bibinfo{author}{\bibfnamefont{T.~G.} \bibnamefont{Steele}},
  \bibnamefont{and} \bibinfo{author}{\bibfnamefont{S.-L.} \bibnamefont{Zhu}},
  \bibinfo{journal}{Phys. Rev. Lett.} \textbf{\bibinfo{volume}{115}},
  \bibinfo{pages}{172001} (\bibinfo{year}{2015}{\natexlab{b}}),
  \eprint{1507.03717}.

\bibitem[{\citenamefont{Roca et~al.}(2015)\citenamefont{Roca, Nieves, and
  Oset}}]{Roca:2015dva}
\bibinfo{author}{\bibfnamefont{L.}~\bibnamefont{Roca}},
  \bibinfo{author}{\bibfnamefont{J.}~\bibnamefont{Nieves}}, \bibnamefont{and}
  \bibinfo{author}{\bibfnamefont{E.}~\bibnamefont{Oset}},
  \bibinfo{journal}{Phys. Rev.} \textbf{\bibinfo{volume}{D92}},
  \bibinfo{pages}{094003} (\bibinfo{year}{2015}), \eprint{1507.04249}.

\bibitem[{\citenamefont{He}(2016)}]{He:2015cea}
\bibinfo{author}{\bibfnamefont{J.}~\bibnamefont{He}}, \bibinfo{journal}{Phys.
  Lett.} \textbf{\bibinfo{volume}{B753}}, \bibinfo{pages}{547}
  (\bibinfo{year}{2016}), \eprint{1507.05200}.

\bibitem[{\citenamefont{Xiao and Meißner}(2015)}]{Xiao:2015fia}
\bibinfo{author}{\bibfnamefont{C.~W.} \bibnamefont{Xiao}} \bibnamefont{and}
  \bibinfo{author}{\bibfnamefont{U.~G.} \bibnamefont{Meißner}},
  \bibinfo{journal}{Phys. Rev.} \textbf{\bibinfo{volume}{D92}},
  \bibinfo{pages}{114002} (\bibinfo{year}{2015}), \eprint{1508.00924}.

\bibitem[{\citenamefont{Aaij et~al.}(2017{\natexlab{a}})}]{Aaij:2017nav}
\bibinfo{author}{\bibfnamefont{R.}~\bibnamefont{Aaij}} \bibnamefont{et~al.}
  (\bibinfo{collaboration}{LHCb}), \bibinfo{journal}{Phys. Rev. Lett.}
  \textbf{\bibinfo{volume}{118}}, \bibinfo{pages}{182001}
  (\bibinfo{year}{2017}{\natexlab{a}}), \eprint{1703.04639}.

\bibitem[{\citenamefont{Yelton et~al.}(2018)}]{Yelton:2017qxg}
\bibinfo{author}{\bibfnamefont{J.}~\bibnamefont{Yelton}} \bibnamefont{et~al.}
  (\bibinfo{collaboration}{Belle}), \bibinfo{journal}{Phys. Rev.}
  \textbf{\bibinfo{volume}{D97}}, \bibinfo{pages}{051102}
  (\bibinfo{year}{2018}), \eprint{1711.07927}.

\bibitem[{\citenamefont{Cheng and Chiang}(2017)}]{Cheng:2017ove}
\bibinfo{author}{\bibfnamefont{H.-Y.} \bibnamefont{Cheng}} \bibnamefont{and}
  \bibinfo{author}{\bibfnamefont{C.-W.} \bibnamefont{Chiang}},
  \bibinfo{journal}{Phys. Rev.} \textbf{\bibinfo{volume}{D95}},
  \bibinfo{pages}{094018} (\bibinfo{year}{2017}), \eprint{1704.00396}.

\bibitem[{\citenamefont{Wang et~al.}(2017{\natexlab{a}})\citenamefont{Wang,
  Xiao, Zhong, and Zhao}}]{Wang:2017hej}
\bibinfo{author}{\bibfnamefont{K.-L.} \bibnamefont{Wang}},
  \bibinfo{author}{\bibfnamefont{L.-Y.} \bibnamefont{Xiao}},
  \bibinfo{author}{\bibfnamefont{X.-H.} \bibnamefont{Zhong}}, \bibnamefont{and}
  \bibinfo{author}{\bibfnamefont{Q.}~\bibnamefont{Zhao}},
  \bibinfo{journal}{Phys. Rev.} \textbf{\bibinfo{volume}{D95}},
  \bibinfo{pages}{116010} (\bibinfo{year}{2017}{\natexlab{a}}),
  \eprint{1703.09130}.

\bibitem[{\citenamefont{Karliner and Rosner}(2017)}]{Karliner:2017kfm}
\bibinfo{author}{\bibfnamefont{M.}~\bibnamefont{Karliner}} \bibnamefont{and}
  \bibinfo{author}{\bibfnamefont{J.~L.} \bibnamefont{Rosner}},
  \bibinfo{journal}{Phys. Rev.} \textbf{\bibinfo{volume}{D95}},
  \bibinfo{pages}{114012} (\bibinfo{year}{2017}), \eprint{1703.07774}.

\bibitem[{\citenamefont{Debastiani et~al.}(2017)\citenamefont{Debastiani, Dias,
  Liang, and Oset}}]{Debastiani:2017ewu}
\bibinfo{author}{\bibfnamefont{V.~R.} \bibnamefont{Debastiani}},
  \bibinfo{author}{\bibfnamefont{J.~M.} \bibnamefont{Dias}},
  \bibinfo{author}{\bibfnamefont{W.~H.} \bibnamefont{Liang}}, \bibnamefont{and}
  \bibinfo{author}{\bibfnamefont{E.}~\bibnamefont{Oset}}
  (\bibinfo{year}{2017}), \eprint{1710.04231}.

\bibitem[{\citenamefont{Chen et~al.}(2018)\citenamefont{Chen, Hosaka, and
  Liu}}]{Chen:2017xat}
\bibinfo{author}{\bibfnamefont{R.}~\bibnamefont{Chen}},
  \bibinfo{author}{\bibfnamefont{A.}~\bibnamefont{Hosaka}}, \bibnamefont{and}
  \bibinfo{author}{\bibfnamefont{X.}~\bibnamefont{Liu}},
  \bibinfo{journal}{Phys. Rev.} \textbf{\bibinfo{volume}{D97}},
  \bibinfo{pages}{036016} (\bibinfo{year}{2018}), \eprint{1711.07650}.

\bibitem[{\citenamefont{Montaña et~al.}(2018)\citenamefont{Montaña, Feijoo,
  and Ramos}}]{Montana:2017kjw}
\bibinfo{author}{\bibfnamefont{G.}~\bibnamefont{Montaña}},
  \bibinfo{author}{\bibfnamefont{A.}~\bibnamefont{Feijoo}}, \bibnamefont{and}
  \bibinfo{author}{\bibfnamefont{A.}~\bibnamefont{Ramos}},
  \bibinfo{journal}{Eur. Phys. J.} \textbf{\bibinfo{volume}{A54}},
  \bibinfo{pages}{64} (\bibinfo{year}{2018}), \eprint{1709.08737}.

\bibitem[{\citenamefont{Nieves et~al.}(2018)\citenamefont{Nieves, Pavao, and
  Tolos}}]{Nieves:2017jjx}
\bibinfo{author}{\bibfnamefont{J.}~\bibnamefont{Nieves}},
  \bibinfo{author}{\bibfnamefont{R.}~\bibnamefont{Pavao}}, \bibnamefont{and}
  \bibinfo{author}{\bibfnamefont{L.}~\bibnamefont{Tolos}},
  \bibinfo{journal}{Eur. Phys. J.} \textbf{\bibinfo{volume}{C78}},
  \bibinfo{pages}{114} (\bibinfo{year}{2018}), \eprint{1712.00327}.

\bibitem[{\citenamefont{Huang et~al.}(2018)\citenamefont{Huang, Xiao, Lü,
  Wang, He, and Geng}}]{Huang:2018wgr}
\bibinfo{author}{\bibfnamefont{Y.}~\bibnamefont{Huang}},
  \bibinfo{author}{\bibfnamefont{C.-j.} \bibnamefont{Xiao}},
  \bibinfo{author}{\bibfnamefont{Q.~F.} \bibnamefont{Lü}},
  \bibinfo{author}{\bibfnamefont{R.}~\bibnamefont{Wang}},
  \bibinfo{author}{\bibfnamefont{J.}~\bibnamefont{He}}, \bibnamefont{and}
  \bibinfo{author}{\bibfnamefont{L.}~\bibnamefont{Geng}}
  (\bibinfo{year}{2018}), \eprint{1801.03598}.

\bibitem[{\citenamefont{Wang et~al.}(2017{\natexlab{b}})\citenamefont{Wang,
  Liu, Kang, and Guo}}]{Wang:2017smo}
\bibinfo{author}{\bibfnamefont{C.}~\bibnamefont{Wang}},
  \bibinfo{author}{\bibfnamefont{L.-L.} \bibnamefont{Liu}},
  \bibinfo{author}{\bibfnamefont{X.-W.} \bibnamefont{Kang}}, \bibnamefont{and}
  \bibinfo{author}{\bibfnamefont{X.-H.} \bibnamefont{Guo}}
  (\bibinfo{year}{2017}{\natexlab{b}}), \eprint{1710.10850}.

\bibitem[{\citenamefont{Liang et~al.}(2017)\citenamefont{Liang, Dias,
  Debastiani, and Oset}}]{Liang:2017ejq}
\bibinfo{author}{\bibfnamefont{W.-H.} \bibnamefont{Liang}},
  \bibinfo{author}{\bibfnamefont{J.~M.} \bibnamefont{Dias}},
  \bibinfo{author}{\bibfnamefont{V.~R.} \bibnamefont{Debastiani}},
  \bibnamefont{and} \bibinfo{author}{\bibfnamefont{E.}~\bibnamefont{Oset}}
  (\bibinfo{year}{2017}), \eprint{1711.10623}.

\bibitem[{\citenamefont{Machleidt et~al.}(1987)\citenamefont{Machleidt,
  Holinde, and Elster}}]{Machleidt:1987hj}
\bibinfo{author}{\bibfnamefont{R.}~\bibnamefont{Machleidt}},
  \bibinfo{author}{\bibfnamefont{K.}~\bibnamefont{Holinde}}, \bibnamefont{and}
  \bibinfo{author}{\bibfnamefont{C.}~\bibnamefont{Elster}},
  \bibinfo{journal}{Phys. Rept.} \textbf{\bibinfo{volume}{149}},
  \bibinfo{pages}{1} (\bibinfo{year}{1987}).

\bibitem[{\citenamefont{Machleidt}(1989)}]{Machleidt:1989tm}
\bibinfo{author}{\bibfnamefont{R.}~\bibnamefont{Machleidt}},
  \bibinfo{journal}{Adv. Nucl. Phys.} \textbf{\bibinfo{volume}{19}},
  \bibinfo{pages}{189} (\bibinfo{year}{1989}).

\bibitem[{\citenamefont{Bedaque and van Kolck}(2002)}]{Bedaque:2002mn}
\bibinfo{author}{\bibfnamefont{P.~F.} \bibnamefont{Bedaque}} \bibnamefont{and}
  \bibinfo{author}{\bibfnamefont{U.}~\bibnamefont{van Kolck}},
  \bibinfo{journal}{Ann. Rev. Nucl. Part. Sci.} \textbf{\bibinfo{volume}{52}},
  \bibinfo{pages}{339} (\bibinfo{year}{2002}), \eprint{nucl-th/0203055}.

\bibitem[{\citenamefont{Machleidt and Entem}(2011)}]{Machleidt:2011zz}
\bibinfo{author}{\bibfnamefont{R.}~\bibnamefont{Machleidt}} \bibnamefont{and}
  \bibinfo{author}{\bibfnamefont{D.~R.} \bibnamefont{Entem}},
  \bibinfo{journal}{Phys. Rept.} \textbf{\bibinfo{volume}{503}},
  \bibinfo{pages}{1} (\bibinfo{year}{2011}), \eprint{1105.2919}.

\bibitem[{\citenamefont{Machleidt}(2017)}]{Machleidt:2017vls}
\bibinfo{author}{\bibfnamefont{R.}~\bibnamefont{Machleidt}},
  \bibinfo{journal}{Int. J. Mod. Phys.} \textbf{\bibinfo{volume}{E26}},
  \bibinfo{pages}{1730005} (\bibinfo{year}{2017}), \eprint{1710.07215}.

\bibitem[{\citenamefont{Machleidt}(2001)}]{Machleidt:2000ge}
\bibinfo{author}{\bibfnamefont{R.}~\bibnamefont{Machleidt}},
  \bibinfo{journal}{Phys. Rev.} \textbf{\bibinfo{volume}{C63}},
  \bibinfo{pages}{024001} (\bibinfo{year}{2001}), \eprint{nucl-th/0006014}.

\bibitem[{\citenamefont{Calle~Cordon and
  Ruiz~Arriola}(2008)}]{CalleCordon:2008cz}
\bibinfo{author}{\bibfnamefont{A.}~\bibnamefont{Calle~Cordon}}
  \bibnamefont{and}
  \bibinfo{author}{\bibfnamefont{E.}~\bibnamefont{Ruiz~Arriola}},
  \bibinfo{journal}{Phys. Rev.} \textbf{\bibinfo{volume}{C78}},
  \bibinfo{pages}{054002} (\bibinfo{year}{2008}), \eprint{0807.2918}.

\bibitem[{\citenamefont{Calle~Cordon and Ruiz~Arriola}(2010)}]{Cordon:2009pj}
\bibinfo{author}{\bibfnamefont{A.}~\bibnamefont{Calle~Cordon}}
  \bibnamefont{and}
  \bibinfo{author}{\bibfnamefont{E.}~\bibnamefont{Ruiz~Arriola}},
  \bibinfo{journal}{Phys. Rev.} \textbf{\bibinfo{volume}{C81}},
  \bibinfo{pages}{044002} (\bibinfo{year}{2010}), \eprint{0905.4933}.

\bibitem[{\citenamefont{Falk and Luke}(1992)}]{Falk:1992cx}
\bibinfo{author}{\bibfnamefont{A.~F.} \bibnamefont{Falk}} \bibnamefont{and}
  \bibinfo{author}{\bibfnamefont{M.~E.} \bibnamefont{Luke}},
  \bibinfo{journal}{Phys. Lett.} \textbf{\bibinfo{volume}{B292}},
  \bibinfo{pages}{119} (\bibinfo{year}{1992}), \eprint{hep-ph/9206241}.

\bibitem[{\citenamefont{Ahmed et~al.}(2001)}]{Ahmed:2001xc}
\bibinfo{author}{\bibfnamefont{S.}~\bibnamefont{Ahmed}} \bibnamefont{et~al.}
  (\bibinfo{collaboration}{CLEO Collaboration}),
  \bibinfo{journal}{Phys.Rev.Lett.} \textbf{\bibinfo{volume}{87}},
  \bibinfo{pages}{251801} (\bibinfo{year}{2001}), \eprint{hep-ex/0108013}.

\bibitem[{\citenamefont{Anastassov et~al.}(2002)}]{Anastassov:2001cw}
\bibinfo{author}{\bibfnamefont{A.}~\bibnamefont{Anastassov}}
  \bibnamefont{et~al.} (\bibinfo{collaboration}{CLEO Collaboration}),
  \bibinfo{journal}{Phys.Rev.} \textbf{\bibinfo{volume}{D65}},
  \bibinfo{pages}{032003} (\bibinfo{year}{2002}), \eprint{hep-ex/0108043}.

\bibitem[{\citenamefont{Gell-Mann and Levy}(1960)}]{GellMann:1960np}
\bibinfo{author}{\bibfnamefont{M.}~\bibnamefont{Gell-Mann}} \bibnamefont{and}
  \bibinfo{author}{\bibfnamefont{M.}~\bibnamefont{Levy}},
  \bibinfo{journal}{Nuovo Cim.} \textbf{\bibinfo{volume}{16}},
  \bibinfo{pages}{705} (\bibinfo{year}{1960}).

\bibitem[{\citenamefont{Sakurai}(1960)}]{Sakurai:1960ju}
\bibinfo{author}{\bibfnamefont{J.~J.} \bibnamefont{Sakurai}},
  \bibinfo{journal}{Annals Phys.} \textbf{\bibinfo{volume}{11}},
  \bibinfo{pages}{1} (\bibinfo{year}{1960}).

\bibitem[{\citenamefont{Kawarabayashi and Suzuki}(1966)}]{Kawarabayashi:1966kd}
\bibinfo{author}{\bibfnamefont{K.}~\bibnamefont{Kawarabayashi}}
  \bibnamefont{and} \bibinfo{author}{\bibfnamefont{M.}~\bibnamefont{Suzuki}},
  \bibinfo{journal}{Phys. Rev. Lett.} \textbf{\bibinfo{volume}{16}},
  \bibinfo{pages}{255} (\bibinfo{year}{1966}).

\bibitem[{\citenamefont{Riazuddin and Fayyazuddin}(1966)}]{Riazuddin:1966sw}
\bibinfo{author}{\bibnamefont{Riazuddin}} \bibnamefont{and}
  \bibinfo{author}{\bibnamefont{Fayyazuddin}}, \bibinfo{journal}{Phys. Rev.}
  \textbf{\bibinfo{volume}{147}}, \bibinfo{pages}{1071} (\bibinfo{year}{1966}).

\bibitem[{\citenamefont{Casalbuoni et~al.}(1993)\citenamefont{Casalbuoni,
  Deandrea, Di~Bartolomeo, Gatto, Feruglio, and Nardulli}}]{Casalbuoni:1992dx}
\bibinfo{author}{\bibfnamefont{R.}~\bibnamefont{Casalbuoni}},
  \bibinfo{author}{\bibfnamefont{A.}~\bibnamefont{Deandrea}},
  \bibinfo{author}{\bibfnamefont{N.}~\bibnamefont{Di~Bartolomeo}},
  \bibinfo{author}{\bibfnamefont{R.}~\bibnamefont{Gatto}},
  \bibinfo{author}{\bibfnamefont{F.}~\bibnamefont{Feruglio}}, \bibnamefont{and}
  \bibinfo{author}{\bibfnamefont{G.}~\bibnamefont{Nardulli}},
  \bibinfo{journal}{Phys. Lett.} \textbf{\bibinfo{volume}{B299}},
  \bibinfo{pages}{139} (\bibinfo{year}{1993}), \eprint{hep-ph/9211248}.

\bibitem[{\citenamefont{Detmold et~al.}(2012)\citenamefont{Detmold, Lin, and
  Meinel}}]{Detmold:2012ge}
\bibinfo{author}{\bibfnamefont{W.}~\bibnamefont{Detmold}},
  \bibinfo{author}{\bibfnamefont{C.~J.~D.} \bibnamefont{Lin}},
  \bibnamefont{and} \bibinfo{author}{\bibfnamefont{S.}~\bibnamefont{Meinel}},
  \bibinfo{journal}{Phys. Rev.} \textbf{\bibinfo{volume}{D85}},
  \bibinfo{pages}{114508} (\bibinfo{year}{2012}), \eprint{1203.3378}.

\bibitem[{\citenamefont{Patrignani et~al.}(2016)}]{Patrignani:2016xqp}
\bibinfo{author}{\bibfnamefont{C.}~\bibnamefont{Patrignani}}
  \bibnamefont{et~al.} (\bibinfo{collaboration}{Particle Data Group}),
  \bibinfo{journal}{Chin. Phys.} \textbf{\bibinfo{volume}{C40}},
  \bibinfo{pages}{100001} (\bibinfo{year}{2016}).

\bibitem[{\citenamefont{Rijken et~al.}(1999)\citenamefont{Rijken, Stoks, and
  Yamamoto}}]{Rijken:1998yy}
\bibinfo{author}{\bibfnamefont{T.~A.} \bibnamefont{Rijken}},
  \bibinfo{author}{\bibfnamefont{V.~G.~J.} \bibnamefont{Stoks}},
  \bibnamefont{and} \bibinfo{author}{\bibfnamefont{Y.}~\bibnamefont{Yamamoto}},
  \bibinfo{journal}{Phys. Rev.} \textbf{\bibinfo{volume}{C59}},
  \bibinfo{pages}{21} (\bibinfo{year}{1999}), \eprint{nucl-th/9807082}.

\bibitem[{\citenamefont{Rijken and
  Yamamoto}(2006{\natexlab{a}})}]{Rijken:2006ep}
\bibinfo{author}{\bibfnamefont{T.~A.} \bibnamefont{Rijken}} \bibnamefont{and}
  \bibinfo{author}{\bibfnamefont{Y.}~\bibnamefont{Yamamoto}},
  \bibinfo{journal}{Phys. Rev.} \textbf{\bibinfo{volume}{C73}},
  \bibinfo{pages}{044008} (\bibinfo{year}{2006}{\natexlab{a}}),
  \eprint{nucl-th/0603042}.

\bibitem[{\citenamefont{Rijken and
  Yamamoto}(2006{\natexlab{b}})}]{Rijken:2006kg}
\bibinfo{author}{\bibfnamefont{T.~A.} \bibnamefont{Rijken}} \bibnamefont{and}
  \bibinfo{author}{\bibfnamefont{Y.}~\bibnamefont{Yamamoto}}
  (\bibinfo{year}{2006}{\natexlab{b}}), \eprint{nucl-th/0608074}.

\bibitem[{\citenamefont{Nagels et~al.}(2014)\citenamefont{Nagels, Rijken, and
  Yamamoto}}]{Nagels:2014qqa}
\bibinfo{author}{\bibfnamefont{M.~M.} \bibnamefont{Nagels}},
  \bibinfo{author}{\bibfnamefont{T.~A.} \bibnamefont{Rijken}},
  \bibnamefont{and} \bibinfo{author}{\bibfnamefont{Y.}~\bibnamefont{Yamamoto}}
  (\bibinfo{year}{2014}), \eprint{1408.4825}.

\bibitem[{\citenamefont{Nagels et~al.}(2015{\natexlab{a}})\citenamefont{Nagels,
  Rijken, and Yamamoto}}]{Nagels:2015lfa}
\bibinfo{author}{\bibfnamefont{M.~M.} \bibnamefont{Nagels}},
  \bibinfo{author}{\bibfnamefont{T.~A.} \bibnamefont{Rijken}},
  \bibnamefont{and} \bibinfo{author}{\bibfnamefont{Y.}~\bibnamefont{Yamamoto}}
  (\bibinfo{year}{2015}{\natexlab{a}}), \eprint{1501.06636}.

\bibitem[{\citenamefont{Nagels et~al.}(2015{\natexlab{b}})\citenamefont{Nagels,
  Rijken, and Yamamoto}}]{Nagels:2015dia}
\bibinfo{author}{\bibfnamefont{M.~M.} \bibnamefont{Nagels}},
  \bibinfo{author}{\bibfnamefont{T.~A.} \bibnamefont{Rijken}},
  \bibnamefont{and} \bibinfo{author}{\bibfnamefont{Y.}~\bibnamefont{Yamamoto}}
  (\bibinfo{year}{2015}{\natexlab{b}}), \eprint{1504.02634}.

\bibitem[{\citenamefont{Geng et~al.}(2008)\citenamefont{Geng, Martin~Camalich,
  Alvarez-Ruso, and Vicente~Vacas}}]{Geng:2008mf}
\bibinfo{author}{\bibfnamefont{L.~S.} \bibnamefont{Geng}},
  \bibinfo{author}{\bibfnamefont{J.}~\bibnamefont{Martin~Camalich}},
  \bibinfo{author}{\bibfnamefont{L.}~\bibnamefont{Alvarez-Ruso}},
  \bibnamefont{and} \bibinfo{author}{\bibfnamefont{M.~J.}
  \bibnamefont{Vicente~Vacas}}, \bibinfo{journal}{Phys. Rev. Lett.}
  \textbf{\bibinfo{volume}{101}}, \bibinfo{pages}{222002}
  (\bibinfo{year}{2008}), \eprint{0805.1419}.

\bibitem[{\citenamefont{Aaij et~al.}(2017{\natexlab{b}})}]{Aaij:2017ueg}
\bibinfo{author}{\bibfnamefont{R.}~\bibnamefont{Aaij}} \bibnamefont{et~al.}
  (\bibinfo{collaboration}{LHCb}) (\bibinfo{year}{2017}{\natexlab{b}}),
  \eprint{1707.01621}.

\bibitem[{\citenamefont{Lewis and Woloshyn}(2009)}]{Lewis:2008fu}
\bibinfo{author}{\bibfnamefont{R.}~\bibnamefont{Lewis}} \bibnamefont{and}
  \bibinfo{author}{\bibfnamefont{R.~M.} \bibnamefont{Woloshyn}},
  \bibinfo{journal}{Phys. Rev.} \textbf{\bibinfo{volume}{D79}},
  \bibinfo{pages}{014502} (\bibinfo{year}{2009}), \eprint{0806.4783}.

\bibitem[{\citenamefont{Savage and Wise}(1990)}]{Savage:1990di}
\bibinfo{author}{\bibfnamefont{M.~J.} \bibnamefont{Savage}} \bibnamefont{and}
  \bibinfo{author}{\bibfnamefont{M.~B.} \bibnamefont{Wise}},
  \bibinfo{journal}{Phys. Lett.} \textbf{\bibinfo{volume}{B248}},
  \bibinfo{pages}{177} (\bibinfo{year}{1990}).

\bibitem[{\citenamefont{Hu and Mehen}(2006)}]{Hu:2005gf}
\bibinfo{author}{\bibfnamefont{J.}~\bibnamefont{Hu}} \bibnamefont{and}
  \bibinfo{author}{\bibfnamefont{T.}~\bibnamefont{Mehen}},
  \bibinfo{journal}{Phys. Rev.} \textbf{\bibinfo{volume}{D73}},
  \bibinfo{pages}{054003} (\bibinfo{year}{2006}), \eprint{hep-ph/0511321}.

\bibitem[{\citenamefont{Guo et~al.}(2013)\citenamefont{Guo, Hidalgo-Duque,
  Nieves, and Valderrama}}]{Guo:2013xga}
\bibinfo{author}{\bibfnamefont{F.-K.} \bibnamefont{Guo}},
  \bibinfo{author}{\bibfnamefont{C.}~\bibnamefont{Hidalgo-Duque}},
  \bibinfo{author}{\bibfnamefont{J.}~\bibnamefont{Nieves}}, \bibnamefont{and}
  \bibinfo{author}{\bibfnamefont{M.~P.} \bibnamefont{Valderrama}},
  \bibinfo{journal}{Phys. Rev.} \textbf{\bibinfo{volume}{D88}},
  \bibinfo{pages}{054014} (\bibinfo{year}{2013}), \eprint{1305.4052}.

\bibitem[{\citenamefont{Choi et~al.}(2011)\citenamefont{Choi, Olsen, Trabelsi,
  Adachi, Aihara et~al.}}]{Choi:2011fc}
\bibinfo{author}{\bibfnamefont{S.-K.} \bibnamefont{Choi}},
  \bibinfo{author}{\bibfnamefont{S.}~\bibnamefont{Olsen}},
  \bibinfo{author}{\bibfnamefont{K.}~\bibnamefont{Trabelsi}},
  \bibinfo{author}{\bibfnamefont{I.}~\bibnamefont{Adachi}},
  \bibinfo{author}{\bibfnamefont{H.}~\bibnamefont{Aihara}},
  \bibnamefont{et~al.}, \bibinfo{journal}{Phys.Rev.}
  \textbf{\bibinfo{volume}{D84}}, \bibinfo{pages}{052004}
  (\bibinfo{year}{2011}), \eprint{1107.0163}.

\bibitem[{\citenamefont{Gamermann and Oset}(2009)}]{Gamermann:2009fv}
\bibinfo{author}{\bibfnamefont{D.}~\bibnamefont{Gamermann}} \bibnamefont{and}
  \bibinfo{author}{\bibfnamefont{E.}~\bibnamefont{Oset}},
  \bibinfo{journal}{Phys.Rev.} \textbf{\bibinfo{volume}{D80}},
  \bibinfo{pages}{014003} (\bibinfo{year}{2009}), \eprint{0905.0402}.

\bibitem[{\citenamefont{Gamermann et~al.}(2010)\citenamefont{Gamermann, Nieves,
  Oset, and Ruiz~Arriola}}]{Gamermann:2009uq}
\bibinfo{author}{\bibfnamefont{D.}~\bibnamefont{Gamermann}},
  \bibinfo{author}{\bibfnamefont{J.}~\bibnamefont{Nieves}},
  \bibinfo{author}{\bibfnamefont{E.}~\bibnamefont{Oset}}, \bibnamefont{and}
  \bibinfo{author}{\bibfnamefont{E.}~\bibnamefont{Ruiz~Arriola}},
  \bibinfo{journal}{Phys. Rev.} \textbf{\bibinfo{volume}{D81}},
  \bibinfo{pages}{014029} (\bibinfo{year}{2010}), \eprint{0911.4407}.

\bibitem[{\citenamefont{Hanhart et~al.}(2011)\citenamefont{Hanhart,
  Kalashnikova, Kudryavtsev, and Nefediev}}]{Hanhart:2011tn}
\bibinfo{author}{\bibfnamefont{C.}~\bibnamefont{Hanhart}},
  \bibinfo{author}{\bibfnamefont{Y.}~\bibnamefont{Kalashnikova}},
  \bibinfo{author}{\bibfnamefont{A.}~\bibnamefont{Kudryavtsev}},
  \bibnamefont{and} \bibinfo{author}{\bibfnamefont{A.}~\bibnamefont{Nefediev}}
  (\bibinfo{year}{2011}), \eprint{1111.6241}.

\bibitem[{\citenamefont{Aaij et~al.}(2014)}]{Aaij:2014ala}
\bibinfo{author}{\bibfnamefont{R.}~\bibnamefont{Aaij}} \bibnamefont{et~al.}
  (\bibinfo{collaboration}{LHCb}), \bibinfo{journal}{Nucl. Phys.}
  \textbf{\bibinfo{volume}{B886}}, \bibinfo{pages}{665} (\bibinfo{year}{2014}),
  \eprint{1404.0275}.

\bibitem[{\citenamefont{Swanson}(2004)}]{Swanson:2004pp}
\bibinfo{author}{\bibfnamefont{E.~S.} \bibnamefont{Swanson}},
  \bibinfo{journal}{Phys. Lett.} \textbf{\bibinfo{volume}{B598}},
  \bibinfo{pages}{197} (\bibinfo{year}{2004}), \eprint{hep-ph/0406080}.

\bibitem[{\citenamefont{Dong et~al.}(2011)\citenamefont{Dong, Faessler,
  Gutsche, and Lyubovitskij}}]{Dong:2009uf}
\bibinfo{author}{\bibfnamefont{Y.}~\bibnamefont{Dong}},
  \bibinfo{author}{\bibfnamefont{A.}~\bibnamefont{Faessler}},
  \bibinfo{author}{\bibfnamefont{T.}~\bibnamefont{Gutsche}}, \bibnamefont{and}
  \bibinfo{author}{\bibfnamefont{V.~E.} \bibnamefont{Lyubovitskij}},
  \bibinfo{journal}{J. Phys.} \textbf{\bibinfo{volume}{G38}},
  \bibinfo{pages}{015001} (\bibinfo{year}{2011}), \eprint{0909.0380}.

\bibitem[{\citenamefont{Guo et~al.}(2015)\citenamefont{Guo, Hanhart,
  Kalashnikova, Meißner, and Nefediev}}]{Guo:2014taa}
\bibinfo{author}{\bibfnamefont{F.-K.} \bibnamefont{Guo}},
  \bibinfo{author}{\bibfnamefont{C.}~\bibnamefont{Hanhart}},
  \bibinfo{author}{\bibfnamefont{{\relax Yu}.~S.} \bibnamefont{Kalashnikova}},
  \bibinfo{author}{\bibfnamefont{U.-G.} \bibnamefont{Meißner}},
  \bibnamefont{and} \bibinfo{author}{\bibfnamefont{A.~V.}
  \bibnamefont{Nefediev}}, \bibinfo{journal}{Phys. Lett.}
  \textbf{\bibinfo{volume}{B742}}, \bibinfo{pages}{394} (\bibinfo{year}{2015}),
  \eprint{1410.6712}.

\bibitem[{\citenamefont{Riska and Brown}(2001)}]{Riska:2000gd}
\bibinfo{author}{\bibfnamefont{D.~O.} \bibnamefont{Riska}} \bibnamefont{and}
  \bibinfo{author}{\bibfnamefont{G.~E.} \bibnamefont{Brown}},
  \bibinfo{journal}{Nucl. Phys.} \textbf{\bibinfo{volume}{A679}},
  \bibinfo{pages}{577} (\bibinfo{year}{2001}), \eprint{nucl-th/0005049}.

\end{thebibliography}

\end{document}